\documentclass[aps,prd,amssymb,twocolumn,superscriptaddress,floatfix,nofootinbib,10pt]{revtex4-2}
\usepackage{graphicx}
\usepackage{dcolumn}
\usepackage{hyperref}
\usepackage[dvipsnames,x11names]{xcolor}
\hypersetup{colorlinks=true,citecolor=Green}
\usepackage{amsmath}
\usepackage{amsfonts}
\usepackage{booktabs}
\usepackage{amssymb}
\usepackage{lineno}
\usepackage{bm}
\usepackage[normalem]{ulem} 
\usepackage{newtxtext,newtxmath}

\usepackage{multirow}
\usepackage{braket}
\usepackage{orcidlink}

\begin{document}


\title{\boldmath Spectroscopic and femtoscopic insights into  vector-baryon interactions in the strangeness $-1$ sector}

\author{P. Encarnación\orcidlink{0009-0005-0749-3885}}
	\email{Pablo.Encarnacion@ific.uv.es}
	\affiliation{Departamento de F\'{i}sica Teórica and IFIC, Centro Mixto Universidad de Valencia-CSIC, Institutos de Investigaci\'{o}n de Paterna, Aptdo. 22085, E-46071 Valencia, Spain}

\author{M. Albaladejo\orcidlink{0000-0001-7340-9235}}
	\email{Miguel.Albaladejo@ific.uv.es}
	\affiliation{Instituto de F\'{i}sica Corpuscular, Centro Mixto Universidad de Valencia-CSIC, Institutos de Investigaci\'{o}n de Paterna, Aptdo. 22085, E-46071 Valencia, Spain}

\author{A. Feijoo\orcidlink{0000-0002-8580-802X}}
        \email{edfeijoo@ific.uv.es}
	\affiliation{Instituto de F\'{i}sica Corpuscular, Centro Mixto Universidad de Valencia-CSIC, Institutos de Investigaci\'{o}n de Paterna, Aptdo. 22085, E-46071 Valencia, Spain}

\author{J. Nieves\orcidlink{0000-0002-2518-4606}}
        \email{jmnieves@ific.uv.es}
	\affiliation{Instituto de F\'{i}sica Corpuscular, Centro Mixto Universidad de Valencia-CSIC, Institutos de Investigaci\'{o}n de Paterna, Aptdo. 22085, E-46071 Valencia, Spain}
 
\date{\today}

\begin{abstract}
We revisit  the strangeness $-1$ sector of the interaction between vector mesons of the $\rho-$nonet  and $1/2^+$ ground state baryons ($VB$), within the unitary coupled-channel hidden gauge  formalism. We adopt a renormalization scheme that induces a reasonable short-distance behavior of the two-hadron wave functions, which is the major limitation of femtoscopy techniques at this time.
We perform an exhaustive spectroscopy study, implementing different improvements and considerations, and  find compatibilities between the poles extracted from the present  approach, and some of the states listed in the Review of Particle Physics. Finally, we predict several correlation functions (CFs) associated with various meson–baryon pairs in this sector, paying special attention to the distinctive and clear signatures produced by the dynamically generated states.  To obtain realistic estimates for the CFs,  we have calculated the production weights using the Thermal-FIST package.  Such studies will shed light into the odd-parity $\Lambda^*$ and $\Sigma^*$ hadron spectra, up to 2 GeV, and will open the strategy to improve effective field theories  by using low-energy constants fitted to femtoscopy data. 
\end{abstract}

\maketitle

\section{Introduction}
\label{sec:introduction}

Despite the existence of kaon factories such as KOTO at J-PARC and DA$\Phi$NE at INFN Frascati, the instability of strange-hadrons makes traditional scattering experiments a challenging task, limiting the experimental access to the strong interaction involving such particles. From the same technical perspective, the situation worsens for heavier hadrons containing charm or bottom quarks, to the point that such experiments are currently unaffordable. In the last few decades, from the pioneering measurements of two light particle correlations~\cite{Pochodzalla:1985zz,Pochodzalla:1985thz,Pochodzalla:1986dqq,Chitwood:1986dfn,Kyanowski:1986gk,Pochodzalla:1987zz,Chen:1987zz} to recent ones carried out by the ALICE Collaboration \cite{ALICE:2022enj,ALICE:2024bhk,ALICE:2019hdt,ALICE:2022yyh,ALICE:2020wvi,ALICE:2021njx,ALICE:2021ovd,ALICE:2019buq,ALICE:2019eol,ALICE:2021cpv,ALICE:2023wjz,ALICE:2023eyl,ALICE:2018ysd,ALICE:2022yyh}, femtoscopy hadron-pair correlation functions (CFs) from high-multiplicity collisions have shown to be a promising source of information on the hadron-hadron interaction.

Clear evidences of the growing interest in these techniques, and in the observables extracted from CFs, are the bidirectional synergies generated in related theoretical studies. On the one hand, hadron-hadron interaction approaches have been tested by reproducing experimental CFs or providing predictions for future and ongoing measurements~\cite{Yan:2025hpa,Albaladejo:2025lhn,Liu:2025oar,Gobel:2025afq,Liu:2025eqw,Abreu:2025jqy,Ikeno:2025kwe,Albaladejo:2025kuv,Encarnacion:2025lyf,Torres-Rincon:2024znb,Etminan:2024nak,Jinno:2024rxw,Feijoo:2024qqg,Etminan:2024uvc,Kamiya:2024diw,Liu:2024nac,Liu:2024uxn,Jinno:2024tjh,Molina:2023oeu,Torres-Rincon:2023qll,Vidana:2023olz,Viviani:2023kxw,Liu:2023wfo,Albaladejo:2023pzq,Liu:2023uly,Kamiya:2022thy,Kamiya:2021hdb,Haidenbauer:2021zvr,Haidenbauer:2020uew,Haidenbauer:2020kwo,Kamiya:2019uiw}. An illustrative case of this situation is the $K^-p$ CF, measured by the ALICE collaboration in $pp$ collisions at $\sqrt{s}=13$ TeV \cite{ALICE:2022yyh}. In that work, it was argued that the chirally motivated models should be revisited based on the discrepancies between the data and the theoretical model \cite{Miyahara:2018onh,Ikeda:2012au} used there to reproduce such a CF. Recently, the authors of Ref.~\cite{Encarnacion:2024jge} aimed at checking this statement, \textit{i.e.} investigating at which level the unitarized chiral approaches are in tension with the femtoscopy data. For that purpose, two chiral models were employed, one at leading order (LO) level \cite{Jido:2003cb,Oset:1997it} and the other up to next-to-leading order (NLO) \cite{Feijoo:2018den}, both properly describing the available $K^-p$ scattering data at low and intermediate energies. The analysis of the CF carried out in Ref.~\cite{Encarnacion:2024jge} showed an agreement within 1$\sigma$ or 2$\sigma$ depending on the set of production weights employed, not only refuting the claim in Ref.\,\cite{ALICE:2022yyh}, but also indirectly demonstrating the compatibility between femtoscopic and scattering data.

 On the other hand, the inverse problem was also addressed either employing model independent approaches \cite{Li:2024tvo,Li:2024tof,Ikeno:2023ojl,Molina:2023jov,Feijoo:2023sfe,Albaladejo:2023wmv} or directly fitting theoretical models to experimental CFs \cite{Feijoo:2024bvn,Mihaylov:2023ahn,Chizzali:2022pjd}. The general scheme employed in the first group of studies starts from a hypothetical measured CF and, by means of the resampling method,  the scattering parameters of the measured hadron pair are determined with certain precision. Alternatively, the fitting procedure, which has been tackled just in a few strangeness sectors, can serve as a powerful strategy to constrain the parameters of the model such as the cutoff used in the regularization of the associated scattering amplitudes, providing information of its interaction range in momentum space, or the low-energy constants (LECs) appearing in the corresponding Lagrangian employed to derive the interaction kernel. Actually, in Ref.\,\cite{Sarti:2023wlg}, the measured $K^- \Lambda$ CF \cite{ALICE:2023wjz} was used as an experimental input to constrain for the first time the LECs of the  $S=-2$ meson-baryon NLO Lagrangian developed in Ref.\,\cite{Feijoo:2023wua}, a sector almost uncharted so far. The resulting unitarized chiral amplitudes provided a more precise pole position and a new distribution of the molecular composition for the $\Xi(1620)$ resonance.

The Koonin–Pratt (KP) CF formalism~\cite{Koonin:1977fh, Pratt:1990zq,Bauer:1992ffu} was recently re-examined in Ref.\,\cite{Albaladejo:2024lam} and compared with an alternative production scheme that would involve the coherent sum of amplitudes instead of the sum of probabilities. The  results of this analysis reinforced the KP approach to theoretically calculate CFs. In principle, the compact and unequivocal structure of the KP formula leads one to think that the femtoscopy data can provide less ambiguous constraints for the theoretical models compared, for instance, to those coming from weak decays. However, the KP formulation depends on the squared modulus of the wave function, which can lead to off-shell effects, rooted in the difference between the ``exact'' wave function and the asymptotic one, which is not correct at short distances. Part of these effects is to give rise, for a sufficiently extended source, to a correction related to the effective range, as already derived in the original formulation by Lednicky and Lyuboshits \cite{Lednicky:1981su,Lednicky:2005tb}, and recently discussed and employed \textit{e.g.} in Refs.\,\cite{Ohnishi:2016elb,Albaladejo:2025kuv,Albaladejo:2025lhn}. However, as pointed out in Ref.\,\cite{Epelbaum:2025aan}, the theoretical computation of the CF using the KP formula is in conflict with the commonly assumed prescription of a universal source due to the off-shell ambiguities discussed above. In this respect, when realistic interactions are taken into account, the possible ambiguities that stem from assuming a universal source model might not significantly affect the analysis, given the current level of measurement accuracy, since these unavoidable ambiguities could fall within the existing systematic and statical errors of the two-body CFs~\cite{Epelbaum:2025aan, Molina:2025lzw}. Indeed,  in Ref.~\cite{Epelbaum:2025aan} it is argued a mild scheme-dependence of $NN$ interactions in chiral effective field theory (EFT), which would result in a moderate model dependence of the two nucleon CFs, if the source term is assumed to be universal. However, the situation is different for three body forces, which provide small but important corrections to the dominant pairwise forces and remain a challenging frontier in nuclear physics and in femtoscopy. 

The theoretical uncertainties on meson-baryon  CFs were estimated to be very small, around 3\% at most for a source radius of around $1\,\text{fm}$ in Ref.~\cite{Molina:2025lzw}, using the coupled-channel Unitarized Chiral Perturbation Theory (UChPT) approach. Another conclusion of this latter analysis is that, once an interaction is chosen, a more correct procedure is to treat the source radius $R$ as an additional parameter of the theory to be fitted.

The sensitivity of the CF to the off-shell behavior of the strong force decreases as the source-radius increases, and  the ambiguities  should become significantly relevant when the extension of the source is large compared to the interaction range, since the large-distance behavior of the scattering wave function is unambiguously determined by the $S-$matrix.

Femtoscopy data undoubtedly contain  extremely  valuable information on  hadron-hadron interactions. Thus, it is an important  theoretical challenge to understand how the off-shell ambiguities of the scattering wave-function are compensated by some interaction-dependent features of the two-hadron emission source, such that both elements provide an acceptable theoretical description of the two-hadron CFs. The latter are genuine observables, since they are obtained from event-number ratios, and their theoretical description should therefore be well defined and free of any ambiguity. Addressing this point is an essential step toward moving beyond a qualitative use of femtoscopy data and entering a new stage of precision, one that would enable the extraction of low-energy hadron--hadron scattering parameters in a theoretically sound and systematically improvable framework. While a complete resolution of this issue is far from trivial, and is currently being addressed by several groups, predicting hadron--hadron CFs using off-shell prescriptions, claimed to be realistic~\cite{Molina:2025lzw}, offers a way forward. These approaches, which go beyond the standard on-shell Lednicky–Lyuboshits formulation \cite{Lednicky:1981su}, may shed light on the role played by off-shell effects and contribute to resolving this fundamental theoretical challenge.

The success of UChPT~\cite{Kaiser:1995cy,Kaiser:1995eg,Oset:1997it}  to describe the low-energy interaction between Goldstone-bosons and the nucleon-octet baryons has been a milestone in the field, being the prediction of the $\Lambda(1405)$ two-pole structure  one of its most notable achievements \cite{Oller:2000fj,Jido:2003cb, Garcia-Recio:2002yxy,Garcia-Recio:2003ejq} [see the dedicated note in the Review of Particle Physics (RPP) \cite{ParticleDataGroup:2024cfk}]. The interaction kernel, typically derived from a lagrangian that respects the relevant symmetries of the system, is the central ingredient of this formalism and is subject to non-perturbative re-summation to restore coupled-channel unitarity. In this context, the hidden gauge symmetry (HGS) formalism \cite{Bando:1984ej,Bando:1987br,Nagahiro:2008cv} provided a sensible framework to incorporate vector-mesons of the $\rho-$nonet into the scheme. One of the seminal studies applying this formalism at LO is the work of Ref.~\cite{Oset:2010tof}, where the dynamics of vector mesons interacting with the baryon octet is investigated. Several dynamically generated states were found, appearing as degenerate $J^P=1/2^-, 3/2^-$ pairs given the limitations of the approach, for different strangeness sectors and are associated with known resonances listed by the RPP. In Ref.~\cite{Garzon:2012np}, other diagrams involving additional coupled channels are taken into account in addition to the driving $t-$channel coming from the local hidden gauge approach. In particular, some box-like diagrams constructed with contact vertices, allowing the coupling of the external vector-baryon pairs to intermediate pseudoscalar-baryon virtual channels, were computed. They contributed only to the $J^P=1/2^-$ sector and broke the original degeneracy. Unfortunately, these effects were rather moderate, resulting primarily in a slight broadening of the dynamically generated states and, in some cases, a small shift of their positions to lower energies. Similarly, in order to break this spin-parity degeneracy, the authors of Refs.\,\cite{Khemchandani:2011et,Khemchandani:2011mf,Khemchandani:2013nma} included additional contact terms, stemming for the local-reduction of  $u-$ and $t-$channel diagrams, which provide the mixing with pseudoscalar-baryon pairs. Another noteworthy approach is that developed in Ref.~\cite{Gamermann:2011mq},  based on a consistent SU(6) flavor-spin symmetry extension of the SU(3) Weinberg-Tomozawa (WT) LO chiral term, within a coupled-channel unitary scheme. This formalism couples all possible combinations of pseudoscalar and vector mesons with ground-state octet and decuplet baryons. Unlike the HGS scheme, this approach naturally breaks the $1/2^--3/2^-$ degeneracy of the resonant states generated by coupling vector mesons and baryons of the nucleon octet.

Recently, the ALICE collaboration measured the $\phi p$ CF in $pp$ collisions at $\sqrt{s}=13$ TeV \cite{ALICE:2021cpv}, and provided the first experimental evidence of an attractive $\phi p$ interaction. In the subsequent analysis, the $\phi p$ scattering length, with an imaginary part compatible with zero within uncertainties, was extracted using the Lednicky-Lyuboshitz approximation~\cite{Lednicky:1981su}, leading the authors to interpret the interaction as being dominated by purely elastic $\phi p$ scattering. In contrast, Ref.\,\cite{Feijoo:2024bvn} presents a theoretical study of the $\phi p$ CF within a unitary extension of the HGS formalism in coupled channels.  This work highlights the crucial role of coupled-channel dynamics in analyzing femtoscopic data. As a direct consequence of the interplay among the coupled channels considered, the extracted scattering length showed significantly better agreement with previous studies and experimental measurements~\cite{Klingl:1997kf,Koike:1996ga,Chang:2007fc,Strakovsky:2020uqs}. A novel aspect of Ref.\,\cite{Feijoo:2024bvn} was the use of the $\phi p$ CF data to constrain the theoretical interaction-model for the first time in that sector. One of the ($J^P=1/2^-,3/2^-$)-degenerate molecular states  contained in the fitted scattering amplitudes appears remarkably close, but above, the $\phi p$ threshold, which is in stark contrast to previous works employing similar approaches where such a pole was found tens of MeV above the threshold~\cite{Oset:2010tof,Garzon:2012np}.

In light of recent and upcoming ALICE measurements of vector–baryon CFs in the $S=0$ sector, it seems natural to extend the theoretical investigation to sectors with nonzero strangeness. Moreover, the rich spectrum of both observed and predicted states near vector–baryon thresholds, particularly those with different strangeness content and whose nature is not yet fully understood, suggests that the corresponding CFs could offer new insights into these states. This is supported by the demonstrated sensitivity of CFs to the existence of near two-hadron threshold states. 

This work aims to provide qualitative reliable theoretical predictions for various CFs, which can be later  confronted with upcoming experimental data. We will focus on  the $S=-1$ sector of the scattering of  vector-mesons of the $\rho-$nonet off $1/2^+$ baryons of the nucleon-octet, which we call $VB$ through this work. We take this opportunity to revisit the unitarized HGS LO formalism derived in Ref.\,\cite{Oset:2010tof}, employing a scale-free dimensional regularization scheme and benefiting from improved experimental knowledge of the resonances, as compared to what was available at the time of its initial application. The obtained poles are compatible with several known resonances listed in the RPP. The assignment of the theoretical states to the $\Sigma(1670)-\Sigma(1750)$ and $\Sigma(1900)-\Sigma(1910)$ resonances appears to be relatively straightforward, whereas the isoscalar states exhibit some ambiguity due to the inherent spin degeneracy of the model. 

The manuscript is organized as follows. In Sec.\,\ref{sec:formalism}, the hidden gauge formalism and the unitarization scheme are introduced, along with the theoretical approach to compute the CF. In Sec.\,\ref{sec:results_T}, the obtained scattering amplitudes are presented, and the found poles are discussed and contrasted to the available RPP information. In Sec.~\ref{sec:results_CF}, we provide novel predictions of the vector-baryon correlation functions in the $S=-1$ sector for future comparison with data. Finally, in Sec.~\ref{sec:conclusions}, the main conclusions of this work are presented.

\section{Formalism}
\label{sec:formalism}

\subsection{Vector-baryon interaction}
The HGS formalism  treats vector meson fields as gauge bosons of a hidden local symmetry while ensuring compatibility with the global chiral symmetry \cite{Bando:1984ej,Bando:1987br,Nagahiro:2008cv}. Within this approach, the three-vector vertex is described by the Lagrangian
\begin{equation}
    \mathcal{L}_{VVV} = ig\left\langle (V^\nu \partial_\mu V_\nu - \partial_\mu V^\nu V_\nu) V^\mu \right\rangle
\end{equation}
where $g=M_V/2f$, $M_V=800$ MeV, $f=93$ MeV and the vector nonet is incorporated through the SU(3) matrix
\begin{equation}
    V_\mu =
    \begin{pmatrix}
    \frac{1}{\sqrt{2}}\rho^0 + \frac{1}{\sqrt{2}}\omega & \rho^+ & K^{*+} \\
    \rho^- & -\frac{1}{\sqrt{2}}\rho^+ + \frac{1}{\sqrt{2}}\omega & K^{*0} \\
    K^{*-} & \bar{K}^{*0} & \phi
    \end{pmatrix}_\mu ,
\end{equation} 
and finally $\langle \cdots \rangle$ stands for the trace in the SU(3) flavor space.

The baryon-baryon-vector vertex $\mathcal{L}_{BBV}$ is given by 
\begin{equation}
    \mathcal{L}_{BBV} = g\left( \langle \bar{B}\gamma_\mu[V^\mu,B] \rangle + \langle \bar{B}\gamma_\mu B \rangle \langle V^\mu \rangle \right) ,
\end{equation}
where $\bar{B}=B^\dagger\gamma^0$ and the baryon-octet  SU(3) matrix-field reads
\begin{equation}
    B =
    \begin{pmatrix}
    \frac{1}{\sqrt{2}}\Sigma^0 + \frac{1}{\sqrt{6}}\Lambda & \Sigma^+ & p \\
    \Sigma^- & -\frac{1}{\sqrt{2}}\Sigma^+ + \frac{1}{\sqrt{6}}\Lambda & n \\
    \Xi^{-} & \Xi^0 & -\frac{2}{\sqrt{6}}\Lambda
    \end{pmatrix} .
\end{equation} 
As shown in Ref.~\cite{Oset:2010tof}, the main contribution to the vector-baryon interaction is obtained from a $t-$channel diagram involving a vector meson acting as mediator, in which the two vertices are derived from the preceding Lagrangians. In the low-energy approximation, upon neglecting the three-momentum of the external vector mesons compared to their characterizing mass $M_V$, the $t-$channel diagram reduces to a contact interaction kernel, which can be expressed as 
\begin{equation}
    {\cal V}_{ij} = -\frac{C_{ij}}{4f^2}  \bar{u}_j(p')\gamma^0 u_i(p)(k^0_i+k^0_j) \vec{\epsilon_i}\cdot\vec{\epsilon_j}^\ast\ ,
    \label{eq:interactionkernel}
\end{equation}
where the indices $i,j$ cover all the possible channels in a given strangeness $S$ sector, $u_i(p)$ is the dimensionless Dirac spinor of the $i$-th channel baryon, $k^0_i$ the energy of the $i$-th channel vector meson and $\vec\epsilon_i$ its polarization vector. Restricting the interaction to its $S-$wave component, the total angular momenta are $J=1/2$ and $J=3/2$, from the coupling of spin 1 and 1/2. Since this interaction kernel is spin-independent in the static limit, both total angular-momentum channels receive identical contributions, resulting in a degenerate interaction. Thus, we take for the $S-$wave projection of the potential 
\begin{equation}
    V_{ij} = -\frac{C_{ij}}{4f^2}  \sqrt{\frac{M_i+E_i}{2M_i}} \sqrt{\frac{M_j+E_j}{2M_j}}\left(2\sqrt{s}-M_i-M_j\right) \ ,
    \label{eq:new_kernel}
\end{equation}
where $M_i$ and $E_i$ are the $i$-th channel baryon mass and energy, respectively, and have included the small $\sqrt{(M_i+E_i)/2M_i}$ terms from the Dirac spinor normalizations, as done in previous studies.

In this work we will concentrate in the $S=-1$ sector. For $I=0$, we consider five coupled channels: $\bar{K}^*N$, $\omega\Lambda$, $\rho\Sigma$, $\phi\Lambda$ and $K^*\Xi$. For $I=1$, we have six channels: $\bar{K}^*N$, $\rho\Lambda$, $\rho\Sigma$, $\omega\Sigma$, $K^*\Xi$ and $\phi\Sigma$. The physical-basis $C_{ij}$ coefficients in Eq.~(\ref{eq:interactionkernel})  are collected in Appendix \ref{app:coefficients}.

To guarantee the unitarity and analyticity of the scattering amplitude, the Bethe-Salpeter equation (BSE) is solved in coupled channels. The interaction kernel (Eq.~(\ref{eq:new_kernel})) and the scattering amplitude involving the intermediate virtual meson-baryon pair are factorized on-shell out of the integral, leaving the BSE as a simple system of equations that in matricial form reads
\begin{equation}
    T = (1-VG)^{-1}V ,
    \label{eq:BS}
\end{equation}
being $G$ a diagonal matrix where each element is the loop function of each vector-baryon pair. According to Eq.~\eqref{eq:interactionkernel}, each ${\cal V}_{kn}$ vertex in the expansion of the BSE loop introduces a $\vec{\epsilon_k}^\ast\cdot\vec{\epsilon_n}$ factor that is also factorized outside of the integral of the loop. Since these intermediate polarizations are tied to the external vector polarizations, one has to sum over all polarizations of the virtual mesons $\left( \sum_{pol} \epsilon_k^\ast\epsilon_n=\delta_{kn}+q_kq_n/M_V^2\right)$. At the end, this procedure gives a correction in the propagator of $\vec{q}^2/3M_V^2$, which can be neglected within the already assumed low-energy approximation. Thus, the resulting vector-baryon loop function is
\begin{equation}
    G_l = i \int \frac{d^4 q_l}{(2\pi)^4} \frac{2M_l}{(P-q_l)^2 - M_l^2 + i\epsilon} \frac{1}{q_l^2-m_l^2+i\epsilon} .
\end{equation}
To render finite this ultraviolet (UV) divergent integral we employ a scale-free dimensional regularization, parameterizing the subtraction constants by means of an UV cutoff $\Lambda$ (see Eqs.~(2.6)-(2.9) of Ref.~\cite{Nieves:2024dcz}),
\begin{widetext}
\begin{eqnarray}
    G_l(s) & = &\frac{1}{4\pi^2} \frac{M_l}{m_l+M_l} \left
(m_l\ln\frac{m_l}{\Lambda + \sqrt{\Lambda^2+m_l^2}}+  M_l\ln\frac{M_l}{\Lambda + \sqrt{\Lambda^2+M_l^2}} \right)
  +  \frac{2M_l}{16\pi^2}  \frac{M_l-m_l}{M_l+m_l}\left( \frac{(M_l+m_l)^2}{s}-1\right) \ln \frac{M_l}{m_l} \nonumber\\
  &+& \frac{M_l\sigma_l}{16\pi^2 s}\Bigg\{\ln\left(s - M_l^2+m_l^2+\sigma_l\right)- \ln \left(-s + M_l^2-m_l^2+\sigma_l\right)+ \ln\left(s + M_l^2-m_l^2+\sigma_l\right) -\ln \left(-s - M_l^2+m_l^2+\sigma_l\right)\Bigg\}\label{eq:loop_hybrid}
\end{eqnarray}
\end{widetext}
where $\sigma_l(s)=(s-(M_l+m_l)^2)^\frac12 (s-(M_l-m_l)^2)^\frac12$ and  $p_l(s)=\sigma_l(s)/(2\sqrt{s})$ is the center-of-mass (CM) momentum of the $l$-th channel.  This renormalization scheme (RS) combines features from two common procedures, one based on the implementation of a sharp UV cutoff and  the other one on dimensional regularization. This RS  preserves the right analytical properties of the loop function, and  at the same time it provides some control on the subtraction constants by means of the UV cutoff, which determines the loop function at the thresholds and thus fixes the numerical values of the subtraction constants for each channel. In this way, the model depends on a single parameter $\Lambda$, chosen to be $830$ MeV in this work.\footnote{This value for the sharp UV cutoff  has been fixed to optimally reproduce the experimental position of the  $I=1$ $\Sigma^*$ resonances found in this work, while the isoscalar $\Lambda^*$ ones are predicted in consequence.}

The presence of unstable particles with large widths, such as the $\rho$ and $K^*$ mesons, requires a convolution of the loop function with their mass distributions, effectively emulating the dressing of the corresponding vector-meson propagator within the loop. Thus, as in Ref.~\cite{Oset:2010tof}, the loop function in Eq.~\eqref{eq:loop_hybrid}  should be replaced by
\begin{eqnarray}
    \widetilde{G}_l(s) & = \displaystyle \frac{1}{N}\int_{(m_l-2\Gamma_l)^2}^{(m_l+2\Gamma_l)^2} dm^2\, \text{Im}\left[ \frac{1}{m^2-m_l^2+im_l\Gamma_l(m)} \right] \nonumber \\
    & \times G_l(s,m^2,M_l^2)\,,
\end{eqnarray}
with the normalization factor $N$
\begin{equation}
    N = \int_{(m_l-2\Gamma_l)^2}^{(m_l+2\Gamma_l)^2} dm^2 \text{Im}\left[ \frac{1}{m^2-m_l^2+im_l\Gamma_l(m)} \right]\,,
\end{equation}
and where the energy-dependent width is
\begin{equation}
    \Gamma_l(m) = \Gamma_l \frac{m_l^2}{m^2} \left( \frac{q(m)}{q(m_l)} \right)^{3} H(m-(m_1+m_2)) \, ,
\end{equation}
which incorporates the $P$-wave decay width $\Gamma_l$,  and the corresponding decay products $m_1$ and $m_2$ for the considered vector meson. In addition, $H(...)$ is the Heaviside function. In the case of the $\rho$, one has $m_1=m_2=m_\pi$ with $\Gamma_\rho=149.77$~MeV, and, for the $K^*$, $ m_1=m_K$, $m_2=m_\pi$ with $\Gamma_{K^*}=48.3$~MeV. Here, $q(x)=\lambda^{1/2}(x^2,m_1^2,m_2^2)/2x$ and $\lambda(a,b,c)$ is the K\"{a}ll\'{e}n function.

\subsection{Poles in the amplitude}

Dynamically generated resonances appear as singularities of the scattering amplitude, obtained through the BSE, on the second Riemann sheet (SRS) at a complex value of $\sqrt{s}$ expressed as $z_0=M_R-{\rm i}\Gamma_R/2$, whose real and imaginary parts correspond to its mass ($M_R$) and half-width ($\Gamma_R/2$). Since these poles arise from the coupling of vector–baryon pairs, they manifest as degenerate $J^P=1/2^-$ and $3/2^-$ states. These resonances can be studied assuming a Breit-Wigner shape for the amplitude,
\begin{equation}
    T_{ij}(z) = \frac{g_ig_j}{z-z_0},
    \label{eq:breitwigner}
\end{equation}
from where the coupling strengths $g_i$ of the resonance to each channel can be determined, as well as its mass and width. At a given complex $\sqrt{s}$, the scattering amplitude in the SRS is obtained modifying the loop in Eq.~\eqref{eq:loop_hybrid} as
\begin{equation}
    G^{II}_l(s)=G_l(s) + i\, 2 M_l \frac{p_l(s)}{4\pi\sqrt{s}}\,,
\end{equation}
for those channels whose threshold is below the real part of $\sqrt{s}$. In this way, the couplings are obtained as follows
\begin{eqnarray}
    g_ig_j = \left[ \frac{\partial}{\partial z}\frac{1}{T_{ij}(z)} \bigg|_{z_0} \right]^{-1} .
    \label{eq:coupling_derivative}
\end{eqnarray}

However, when the mass distributions for the $\rho$ and $K^*$ vector mesons are taken into account, the channel thresholds, and thus the transitions between Riemann sheets, become diffuse. For instance, in the present case, the $\Sigma^*$ states lie sufficiently close to the $\bar{K}^*N$ and $\rho\Sigma$ thresholds and, consequently, the shape of the amplitude line is strongly affected by the convolution. More precisely, the real part of the loop function could be softened because of the vector-meson width to the extent that there is no crossing\footnote{This can be understood from the denominator of Eq.~\eqref{eq:BS} and implementing  the replacement $G\to\widetilde{G}$.} between the loop function $\widetilde{G}$
and the inverse of the kernel $V^{-1}$, thus causing either a dilution of the corresponding amplitude or even the disappearance of the pole. Instead, following Ref.\,\cite{Oset:2010tof}, a good approximation is to study these poles in the real axis, where the amplitude reaches its maximum at $\sqrt{s}=M_R$. Therefore, up to a global sign, one can obtain $g_i$ that corresponds to the channel that couples the most to the resonance from Eq.~\eqref{eq:breitwigner},
\begin{equation}
    g_i^2 = i\frac{\Gamma_R}{2}T_{ii}(\sqrt{s}=M_R) ,
    \label{eq:coupling_real}
\end{equation}
while the rest of the couplings are obtained from the previous one as
\begin{equation}
    g_j=g_i\frac{T_{ij}(\sqrt{s}=M_R)}{T_{ii}(\sqrt{s}=M_R)} .
    \label{eq:coupling_real_ij}
\end{equation}
and $\Gamma_R$ is estimated from the peak seen in the absolute value squared of the elastic channel $i$, which couples most to the resonance.

\subsection{Correlation function}
\label{sec:correlationfunction}

The meson-baryon CF in our multichannel scheme for the measured final channel $i$ is given by the generalized KP formula \cite{Lisa:2005dd,Haidenbauer:2018jvl,Vidana:2023olz},
\begin{equation}
    C_i(p) = \sum_j w_j \int d^3 r S_j(r) |\psi_{ji}(p,r)|^2
    \label{eq:kooninpratt}
\end{equation}
where $p$ is the meson-baryon pair relative CM momentum. The summation includes all possible intermediate $j$-th channels that can couple to the measured $i$-th one. Each of these contributions is weighted by  a factor, $w_j$, which accounts for the relative production strength of each channel with respect to the measured pair $i$. The production weights used in this work\footnote{They have been calculated for high-multiplicity events (average charged particle multiplicity at mid-rapidity $<dN_{ch}/d\eta>=30$), employing the production yields obtained through the $\gamma_S$CSM model, implemented in the Thermal-FIST package \cite{Vovchenko:2019pjl} using the relative momentum distributions extracted from the CBW model \cite{Schnedermann:1993ws}. We follow the VLC method explained in the Appendix~A of Ref.~\cite{Encarnacion:2024jge}.} are collected in Table~\ref{tab:weights}.  In addition, in Eq.~\eqref{eq:kooninpratt}, $S_j(r)$ is the so-called source function, which stands for the probability distribution of emitting a $j$-pair at a relative distance $r$. The ALICE collaboration \cite{2020135849,ALICE:2023sjd} showed that, in $pp$ collisions, this source has two components: a gaussian core common to all particles and a non-gaussian component coming from strongly decaying resonances into the particles forming the studied pair. The source function is typically parametrized as an effective spherical Gaussian $S_j(r)=(4\pi R_j^2)^{-3/2}\exp(-r^2/4R_j^2)$, where its size $R_j$ is, in general,  channel dependent due to the different decay contributions to the particles of the pair. 

To obtain the most realistic estimates possible of the CFs, we use information of the source-radii from previous works. Given that in Refs.~\cite{ALICE:2022yyh} and \cite{ALICE:2021cpv}, identical values for $R_{\bar{K}N}=1.08$ fm and $R_{\phi N}=1.08$ fm (where the $\phi$ yields were considered to be 100\% primordial) are obtained, we hypothesize  that the decays into nucleons dominate the source size, and  thus take $R_{\bar{K}^*N}=1.08$ fm. On the other hand, considering the $\Lambda$ and $\Sigma$ to be equivalent, as done in \cite{ALICE:2022yyh},  then we have assumed that the $\rho\Lambda,\omega\Lambda,\phi\Lambda,\rho\Sigma,\omega\Sigma,\phi\Sigma$ sources are equivalent to the one reported in the experimental work of Ref.~\cite{ALICE:2023wjz} for the emission of the $\bar{K}\Lambda$ pair.\footnote{According to Refs.~\cite{ALICE:2022yyh} and \cite{ALICE:2020ibs}, 52\%(36\%) of the total produced $\bar{K}$($\Lambda$) are primordial, and since the decay contributions to  $\bar{K}$ mesons are shorter-lived than those to  $\Lambda$ baryons,  we believe a reasonable approximation to consider that the $\bar{K}\Lambda$ radius is dominated by the emission of the $\Lambda$ hyperons.} In that work, a double gaussian parametrization was used, but we find it to be approximately equivalent to a single gaussian of radius $1.28$ fm, hence we use $R_{V\Lambda}=R_{V\Sigma}=1.28$ fm. For the heavier $K^*\Xi$ channels we simply use $R_{K^*\Xi}=1.0$ since the contributions of these channels to the studied CFs in this work are barely relevant, as will be shown later. In the uncertainty propagation we associate a 10\% error to all these radii, similarly to the uncertainties given in the aforementioned ALICE works.

\begin{table*}
    \centering
    \begin{tabular}{|c|c|c|c||c|c|c|c|} \hline
        Channel ($Q=+1$) & $\bar{K}^{*0}p$ CF & $\rho^+\Lambda$ CF        & $\rho^0\Sigma^+$ CF          & Channel ($Q=0$)  & $\omega\Lambda$ CF & $\rho^0\Sigma^0$ CF    & $\phi\Lambda$ CF \\  [1mm] \hline
    $\bar{K}^{*0}p$      & $1$                & $1.49\pm0.11$             & $2.15\pm0.17$                & $\bar{K}^{*0}n$  & $1.52\pm0.12$      & $2.17\pm0.18$          & $5.91\pm0.53$   \\ [1.5mm]
    $\rho^+\Lambda$      & $0.25\pm 0.02$     & $1$                       & $1.69\pm0.13$                & $K^{*-}p$        & $1.58\pm0.12$      & $2.18\pm0.16$          & $5.55\pm0.51$   \\ [1.5mm]
    $\rho^0\Sigma^+$     & $0$                & $0.109\pm0.008$           & $1$                          & $\rho^0\Lambda$  & $1.09\pm0.09$      & $1.68\pm0.13$          & $4.49\pm0.42$   \\ [1.5mm]
    $\rho^+\Sigma^0$     & $0$                & $0.080\pm0.006$           & $0.90\pm0.07$                & $\omega\Lambda$  & $1$                & $1.65\pm0.11$          & $4.56\pm0.39$   \\ [1.5mm]
    $\omega\Sigma^+$     & $0$                & $0.028\pm0.002$           & $0.78\pm0.06$                & $\rho^-\Sigma^+$ & $0.197\pm0.016$    & $1.03\pm0.08$          & $3.45\pm0.28$   \\ [1.5mm]
    $\phi\Sigma^+$       & $0$                & $0$                       & $0$                          & $\rho^0\Sigma^0$ & $0.172\pm0.013$    & $1$                    & $3.48\pm0.31$   \\ [1.5mm]
    $K^{*+}\Xi^0$        & $0$                & $0$                       & $0$                          & $\rho^+\Sigma^-$ & $0.127\pm0.010$    & $0.88\pm0.07$          & $3.27\pm0.27$   \\ [1.5mm]
        - & - & - & -                                                                                    & $\omega\Sigma^0$ & $0.080\pm0.006$    & $0.78\pm0.06$          & $3.43\pm0.29$   \\ [1.5mm]
        - & - & - & -                                                                                    & $\phi\Lambda$    & $0$                & $0$                    & $1$             \\ [1.5mm]
        - & - & - & -                                                                                    & $K^{*0}\Xi^0$    & $0$                & $0$                    & $0.046\pm0.004$ \\ [1.5mm]
        - & - & - & -                                                                                    & $\phi\Sigma^0$   & $0$                & $0$                    & $0.020\pm0.002$ \\ [1.5mm]
        - & - & - & -                                                                                    & $K^{*+}\Xi^-$    & $0$                & $0$                    & $0.001\pm0.000$ \\   [1.5mm] \hline
    \end{tabular}
    \caption{Production weights for the $\bar{K}^{*0}p$, $\rho^+\Lambda$, $\rho^0\Sigma^+$, $\omega\Lambda$, $\rho^0\Sigma^0$ and $\phi\Lambda$ channels used to calculate the CFs in this work. These values are obtained following the VLC method (see a detailed explanation of the methodology in Appendix~A of Ref.~\cite{Encarnacion:2024jge}).}
    \label{tab:weights}
\end{table*}

The last element to be computed in Eq.~\eqref{eq:kooninpratt} is the relative wave function, $\psi_{ji}(p,r)$, for the transition of the intermediate channel $j$ to the measured one $i$. This wave function is obtained from the half off-shell scattering amplitude, for which only its $S-$partial wave is non-zero. The meson-baryon $S-$wave function reads
\begin{eqnarray}
    \psi_{ji}(p,r) = \delta_{ji}j_0(pr) + \int \frac{d^3q}{(2\pi)^3} \frac{1}{2\omega_j(q)}\frac{M_j}{E_j(q)} \nonumber \\
    \times \frac{T_{ji}(\sqrt{s},p,q)\ j_0(qr)}{\sqrt{s}-\omega_j(q)-E_j(q)+i\epsilon} \ ,
    \label{eq:wf_ampli} 
\end{eqnarray}
where $\omega_j(q)=(q^2+m_j^2)^{1/2}$ and $E_j(q)=(q^2+M_j^2)^{1/2}$ are the energies of the vector meson and the baryon, respectively, in the $j$-th channel, and $j_0(x)$ is the spherical Bessel function of the first kind. Following Ref.~\cite{Vidana:2023olz}, we assume a half off-shell $T-$matrix expressed as
\begin{equation}
T_{ji}(\sqrt{s},p,q))= H(\Lambda-p)T_{ji}(\sqrt{s})H(\Lambda-q)\, ,
\label{eq:hos_scat}
\end{equation}
where $T_{ji}(\sqrt{s})$ is the corresponding on-shell scattering matrix element and the $H(\Lambda-k)$ denotes the pertinent step functions, with $\Lambda$ the UV cutoff used to renormalize the on-shell scattering amplitude~\cite{Oset:2010tof}. The prescription introduced in Eq.~\eqref{eq:hos_scat}, besides rendering the integral over \( q \) in Eq.~\eqref{eq:wf_ampli} finite, also allows an on-shell factorization of the \( T_{ji}(\sqrt{s}) \) automatically.

Note that Eq.~\eqref{eq:wf_ampli}  involves the vector-baryon loop functions, but these do not need to be dressed to take into account the $\rho$ and $K^*$ widths, since they are included in the scattering amplitude and therefore in the wave function. Furthermore, as argued in Ref.~\cite{Feijoo:2024bvn}, the finite width effects of $\rho$ and $K^*$ mesons are incorporated into the production weights when the  yields are estimated within a thermal model. In the study of Ref.~\cite{Feijoo:2024bvn}, the influence of the $\rho$ width on the calculated relative momentum ($p$) on the paired proton was also investigated, and it was found that the effect of such a width was negligible compared to the zero width case. In fact, the shift in the relative momentum $p$ caused by the finite meson widths lies well below the current experimental resolution.

\subsection{Scattering parameters}
\label{sec:sp}
In our normalization, the single-channel scattering matrix $T$ obtained from Eq.~\eqref{eq:BS} is related to the usual Quantum Mechanics amplitude $f^{\text{QM}}$ by:
\begin{align}
     T & = -\frac{8\pi\sqrt{s}}{2M}f^{\text{QM}}\hspace{0.2cm}  ; \nonumber \\
     (f^{\text{QM}})^{-1} & \simeq -\frac{1}{a_{0}}+\frac{1}{2}r_{0}p^2 + {\cal O}(p^4)- i p\,,
    \label{eq:QMamplitude}
\end{align}
being $p$ the CM momentum of the system. We can characterize the coupled-channel amplitudes near threshold computing the parameters
\begin{subequations}\label{eq:a0r0}
\begin{align}
    \frac{1}{a_{0,i}} &= \frac{8\pi\sqrt{s}}{2M_i} \frac{1}{T_{ii}}\bigg|_{s=s_i}\,, \\
    r_{0,i} &= \frac{1}{\mu_i}\frac{\partial}{\partial\sqrt{s}} \left[ -\frac{8\pi\sqrt{s}}{2M_i}\left(\frac{1}{T_{ii}}+{i\,\rm Im}[\widetilde G_i]\right) \right]_{s=s_i}\,,
\end{align}
\end{subequations}
being $\mu_i$ the reduced mass of the $i$-th channel and $s_i=(M_i+m_i)^2$ the threshold energy squared. Note that the factor  ${-8\pi\sqrt{s}}\,{\rm Im}[\widetilde G_i]/(2M_i)$ becomes just $p_i$ (center of mass momentum of the $i$-th channel), when the width  of the vector mesons is not taken into account.

\section{Results}

We now present the  results obtained in this work. First, we examine both isoscalar ($I=0$ )  and isovector ($I=1)$  $VB$ poles, in the $S=-1$ sector, generated in the scattering amplitudes within the HGS formalism, and we compare them with their possible counterparts in the RPP~\cite{ParticleDataGroup:2024cfk}. We discuss the SU(3) limit of the spectrum and calculate the scattering parameters, as defined in subsec~\ref{sec:sp}. We then give  predictions for different measurable CFs of physical channels, studying the contribution of each inelastic transition to the corresponding femtoscopy observable. Finally, the scattering parameters for the studied vector-baryon pairs are also provided.

\begin{figure}
    \centering
    \includegraphics[width=1\columnwidth]{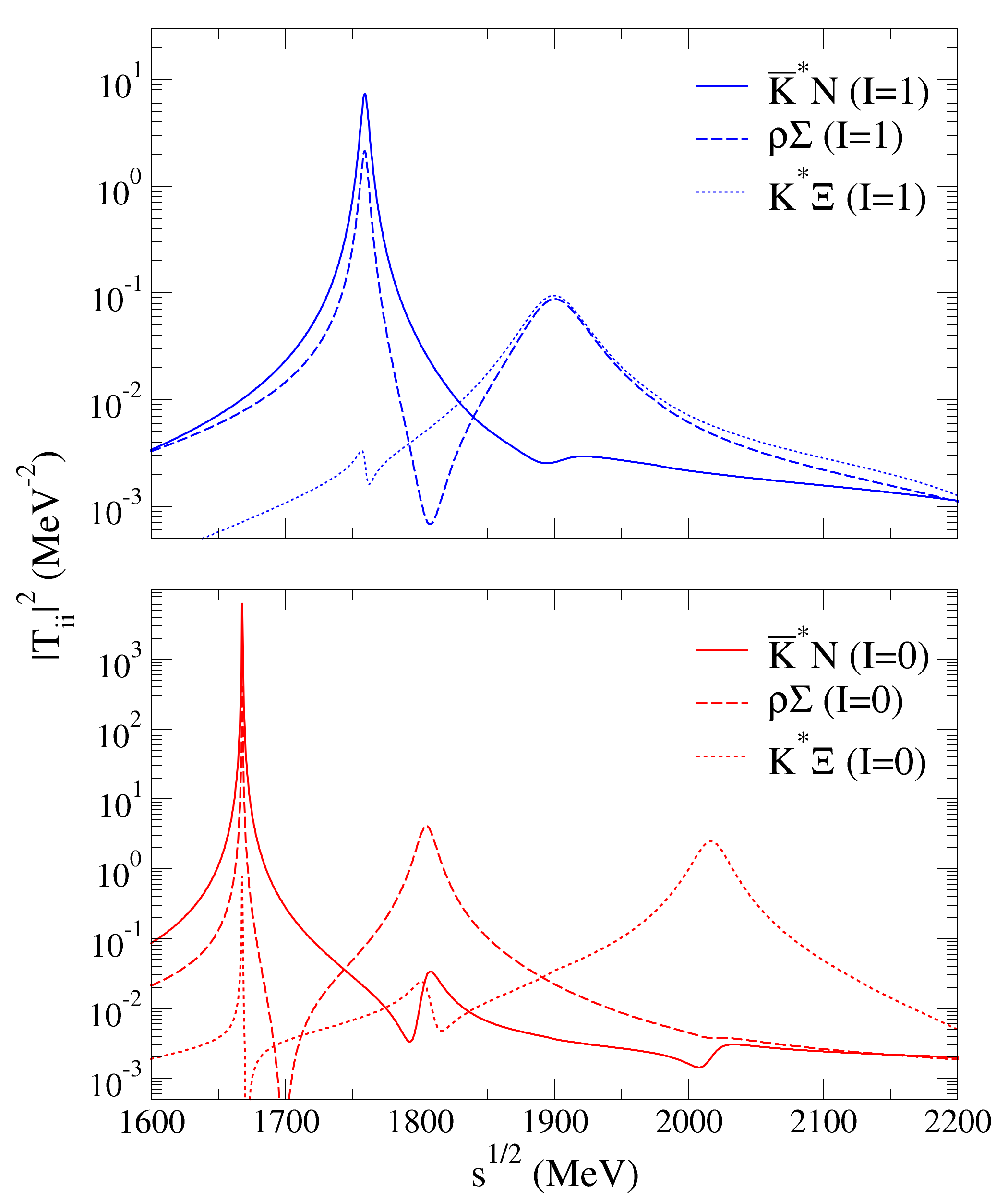}
    \caption{Absolute value squared of the elastic $\bar K^*N$, $\rho\Sigma$ and $K^*\Xi$  scattering amplitudes for both  $I=1$ (top) and $I=0$ (bottom) sectors. Results have been obtained using a sharp UV cutoff $\Lambda=830\,\text{MeV}$.  }
    \label{fig:isospin_amplitude}
\end{figure}

\subsection{Dynamically generated resonances}
\label{sec:results_T}

\subsubsection{Amplitude Signatures }
In Fig.~\ref{fig:isospin_amplitude}, we display the most relevant elastic amplitudes ($|T_{ii}|^2$) as a function of $\sqrt{s}$ in the $I=0$ and $I=1$ sectors. One should bear in mind that, within the employed LO formalism, the dynamically generated states are degenerate since the $J^P=1/2^-$ and $3/2^-$ amplitudes are identical. In $I=0$ one clearly distinguishes three $\Lambda$ states. Focusing first on the elastic $|T_{\bar{K}^*N}|^2$ (solid-red line in the bottom panel of Fig.~\ref{fig:isospin_amplitude}), a clear peak appears around $1670$ MeV, which also manifests in the other elastic squared amplitudes as an evidence of the intricate molecular composition of this state. This behavior can be better understood by inspecting the first column of Table~\ref{tab:couplings_I0}, where, although the bound state couples strongly to the $\bar{K}^*N$ channel, other sizable couplings $|g_{i}|$ also emerge as a result of the coupled-channel dynamics. The next $I=0$ structure shows up as a bump at $1800$ MeV in $\rho\Sigma\to\rho\Sigma$ elastic transition (see dashed-red line in the bottom panel of Fig.~\ref{fig:isospin_amplitude}),  but is barely visible in the other channels, which is in accordance with the corresponding couplings compiled in the second column of Table~\ref{tab:couplings_I0}. In this case, despite lying below the lowest threshold, the resonance acquires a finite width due to the convolution with the $\rho$ mass distribution, whose high-mass tail contributes above threshold, thereby opening up phase space for the $\pi\pi\Sigma$ decay. The third state is clearly distinguishable only in the $K^*\Xi$ channel around $2000$ MeV (dotted-red line in the bottom panel of Fig.~\ref{fig:isospin_amplitude}). The coupling pattern revealed in the last column of Table~\ref{tab:couplings_I0} supports an interpretation of this state  mostly as a quasi-bound $K^*\Xi$ molecule.

In the isovector sector, signatures for two $\Sigma^*$ states can be identified in the top panel of  Fig.~\ref{fig:isospin_amplitude}. The first peak appears at $1750$ MeV  in the $\bar{K}^*N$ elastic amplitude (solid blue line). The state also couples strongly to the $\rho\Sigma$ channel, leaving a clear signal in its elastic amplitude (dashed blue curve). As occurred previously for the $\Lambda(1805^*)$, the finite width of this state arises from the convolution with the mass distribution of the $K^*$, which high-mass tail accounts for the decay into the $\bar K \pi N$ three body channel. Additionally, a second $\Sigma^*$ resonance is clearly visible in the $\rho\Sigma$ and $K^*\Xi$ elastic amplitudes, exhibiting a clear structure with comparable strength in both cases, as seen in the dotted and dashed lines in the top panel of Fig.~\ref{fig:isospin_amplitude}. The effect on the $\bar K^*N$ is much weaker. This is corroborated by  the coupling values of  this $\Sigma(1900^*)$ resonance to all these channels in Table~\ref{tab:couplings_I1}. 
\begin{table}
    \resizebox{.95\columnwidth}{!}{
    \centering
    \begin{tabular}{c|cc|cc|cc} \hline
         \multirow{2}{*}{$I=0$} & \multicolumn{2}{c|}{$1668-0i$} & \multicolumn{2}{c|}{$1805-20.14i$} & \multicolumn{2}{c}{$2018-32.90i$} \\
         & $g_i$ & $|g_i|$ & $g_i$ & $|g_i|$ & $g_i$ & $|g_i|$ \\  \hline
$\bar{K}^*N$ (1833) & $4.27$ & $4.27$ & $0.52+0.12i$ & $0.54$ & $0.37+0.30i$ & $0.48$ \\
$\omega\Lambda$ (1898) & $1.26$ & $1.26$ & $0.38+0.10i$ & $0.39$ & $-0.75-0.13i$ & $0.76$ \\
$\rho\Sigma$ (1964) & $-2.14$ & $2.14$ & $1.90+0.59i$ & $1.99$ & $-0.16-0.14i$ & $0.21$ \\
$\phi\Lambda$ (2135) & $-1.73$ & $1.73$ & $-0.52-0.14i$ & $0.54$ & $1.03+0.18i$ & $1.05$ \\
$K^*\Xi$ (2212) & $-0.44$ & $0.44$ & $0.41+0.12i$ & $0.43$ & $4.63-0.31i$ & $4.64$ \\
        \hline
    \end{tabular}
    }
    \caption{Positions ($M_R-i\Gamma_R/2)$ (in MeV units) and coupling strengths of the three dynamically generated $\Lambda$ resonances discussed in Fig.~\ref{fig:isospin_amplitude}.  In the first column, the numbers in brackets stand for the thresholds (in MeV) of each channel.}
    \label{tab:couplings_I0}
\end{table}

\begin{table}
    \centering
    \begin{tabular}{c|cc|cc} \hline
         \multirow{2}{*}{$I=1$} & \multicolumn{2}{c|}{$1759- 3.06i$} & \multicolumn{2}{c}{$1900-24.72i$}  \\
         & $g_i$ & $|g_i|$ & $g_i$ & $|g_i|$  \\  \hline
$\bar{K}^*N$ (1833) & $\hphantom{+}2.89-0.08i$ & $2.89$ & $\hphantom{+}0.25-0.57i$ & $0.62$  \\
$\rho\Lambda$ (1887) & $-1.55+0.09i$ & $1.55$ & $-0.93+0.55i$ & $1.08$ \\
$\rho\Sigma$ (1964) & $-2.11+0.08i$ & $2.11$ & $\hphantom{+}2.70-0.12i$ & $2.70$  \\
$\omega\Sigma$ (1976) & $-0.84+0.05i$ & $0.84$ & $-0.51+0.30i$ & $0.59$  \\
$K^*\Xi$ (2212) & $\hphantom{+}0.25+0.01i$ & $0.25$ & $\hphantom{+}2.62-0.54i$ & $2.68$  \\
$\phi\Sigma$ (2213) & $\hphantom{+}1.16-0.07i$ & $1.16$ & $\hphantom{+}0.70-0.41i$ & $0.81$  \\
        \hline
    \end{tabular}
    \caption{Same as Table~\ref{tab:couplings_I0}, but for the two $\Sigma$ resonances found in this work.  The states are analysed in the real axis, as discussed  in the context of Eqs.~\eqref{eq:coupling_real} and \eqref{eq:coupling_real_ij}.}
    \label{tab:couplings_I1}
\end{table}

\begin{figure}
    \centering
    \includegraphics[width=\columnwidth]{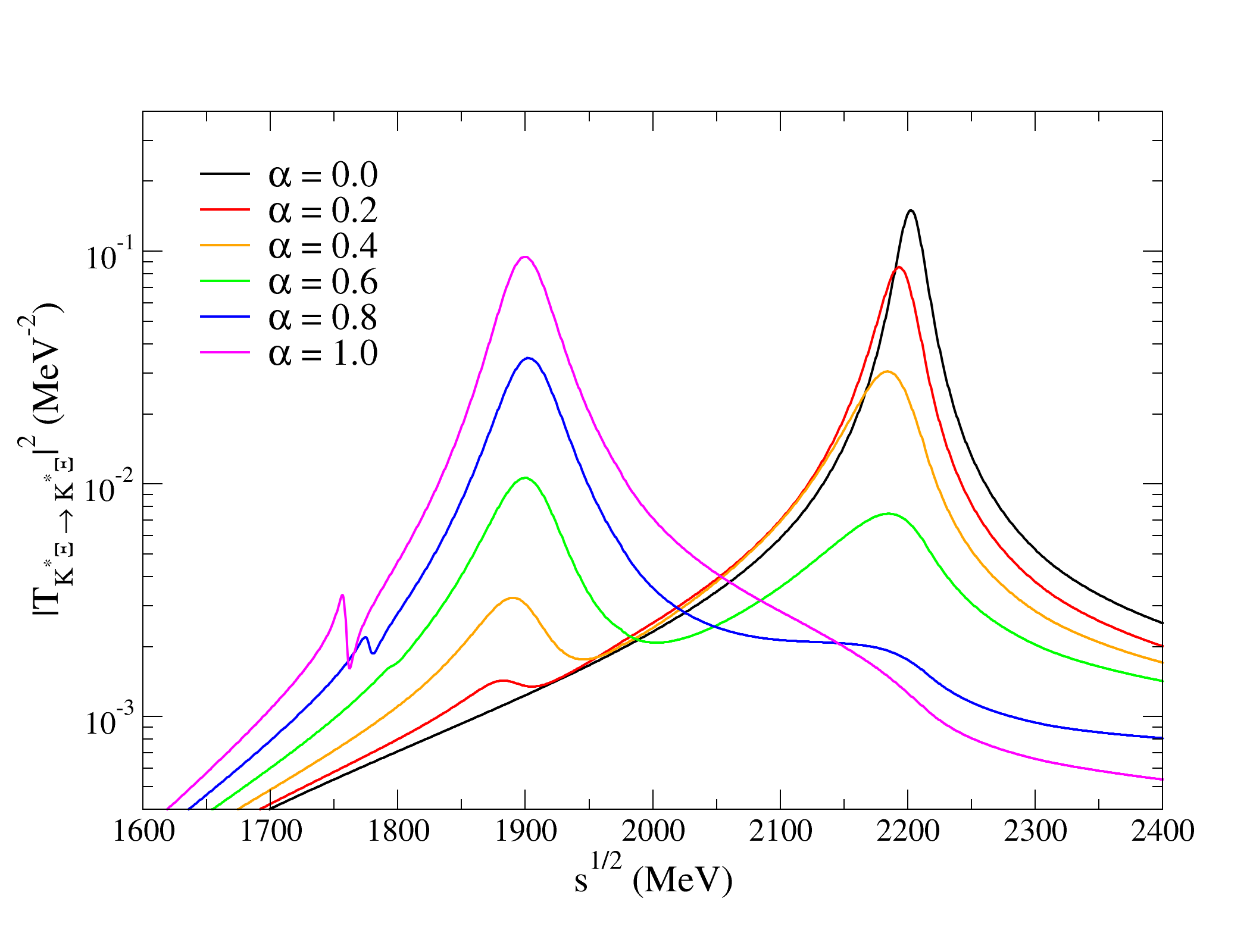}
    \caption{Absolute value squared of the elastic  $K^*\Xi$ amplitude in the $I=1$ sector 
    for different values of the parameter $\alpha$ regulating the strength of the coupled channels (see text for details).}
    \label{fig:coupling_amplitude}
\end{figure}

The Clebsch–Gordan-type coefficients $C_{ij}$ preceding the interaction kernels, derived from the employed Lagrangians, not only encode the symmetries of the system, but the diagonal terms also carry information about the attractive or repulsive character of the interaction, depending on their sign. An attractive interaction\footnote{Note that in Eq.~\eqref{eq:interactionkernel} we have adopted a sign convention in which positive values of the diagonal matrix elements $C_{ii}$ indicate an attractive interaction, and negative values indicate a repulsive one.} is the {\it sine qua non} condition for the formation of a hadron–hadron bound state.  In addition to the resonances extracted from the physical amplitudes in Fig.~\ref{fig:isospin_amplitude}, a possible isovector resonance might be expected close to the $K^*\Xi$ threshold, since the corresponding diagonal \(C_{ii}\) coefficient is attractive. Despite $C_{K^*\Xi,K^*\Xi} = 1$ (see Table 9 of Ref.\,\cite{Oset:2010tof}), such a state does not appear. In view of this, we studied the issue in detail. In Fig.~\ref{fig:coupling_amplitude}, the absolute value squared of the elastic $I=1$ $K^*\Xi$ amplitude is displayed in terms of a parameter $\alpha$, which controls the inelastic couplings in Eq.~\eqref{eq:interactionkernel}. That is, we replace $C_{ij}$ by $\alpha C_{ij}$ with $i\neq j$. In the uncoupled channel limit ($\alpha=0$), we observe  that the signature of a  new state appears around 2200 MeV just below the $K^*\Xi$ threshold, which gradually dissolves as $\alpha$ increases, simultaneously favoring the emergence of the other two $\Sigma^*$ states discussed in Fig.~\ref{fig:isospin_amplitude}. This fact provides clear evidence of the complexity of the coupled-channel dynamics within the HGS formalism. 

This five-state (three $\Lambda^*$ and two $\Sigma^*$) picture found in the $S=-1$ sector of the scattering  of the vector-mesons of the $\rho-$nonet off $1/2^+$ baryons of the nucleon-octet can be understood from the point of view of group theory. The SU(3) reduction into irreducible representations (irreps) reads~\cite{deSwart:1963pdg} $(8\oplus1)\otimes8=1\oplus8\oplus8\oplus10\oplus10^*\oplus27\oplus8$. Note that in the SU(3) basis $\ket{\phi;I;I_3;Y}$, where $\phi$ denotes the corresponding irrep and $Y=B+S$ is the hypercharge, the interaction should be diagonal, except perhaps in the sub-space of the octets, and thus the eigenvalues of the matrix $C_{ij}$ indicate the attractive or repulsive character of each irrep. Looking at the eigenvalues ($\lambda$) of the $I=0$, $I=1$ and $I=2$ coefficients once diagonalized (see Tables 8, 9 and 10 in Ref.~\cite{Oset:2010tof}), one immediately identifies a very bound singlet ($\lambda=6$), two bound octets ($\lambda=3$), a repulsive 27-plet ($\lambda=-2$) and non-interacting $10$, $10^*$ and $8$-plets ($\lambda=0$).  These eigenvalues support the existence of three $\Lambda^*$ and two $\Sigma^*$ states in the $S=-1$ sector, since the octet contains both $\Lambda$ and $\Sigma$ hyperons, while the singlet is formed by a baryon with $\Lambda$ quantum-numbers. On top of that, we have the accidental spin-parity $(1/2^--3/2^-)$ degeneracy present in the LO HGS vector meson-baryon interaction.  

The dynamics of the $I=1$ sector analyzed in Fig.~\ref{fig:coupling_amplitude} is explained by the evolution of these eigenvalues with the parameter $\alpha$ introduced above. For $\alpha<1$ there are four positive (attractive), one negative (repulsive), and two zero (non-interacting) eigenvalues. The attractive irreps are the singlet, which does not contain $\Sigma-$like states, and the two octets and the anti-decuplet. As $\alpha$ approaches one, the positive eigenvalue of the  $10^*$ irrep smoothly approaches zero, and the corresponding resonance disappears, in accordance with magenta curve in  Fig.~\ref{fig:coupling_amplitude}. The role played by the coupled channels prevents this state from being formed, reflecting the importance of considering them.

\subsubsection{SU(3) limit }
\begin{figure}
    \centering
    \includegraphics[width=\columnwidth]{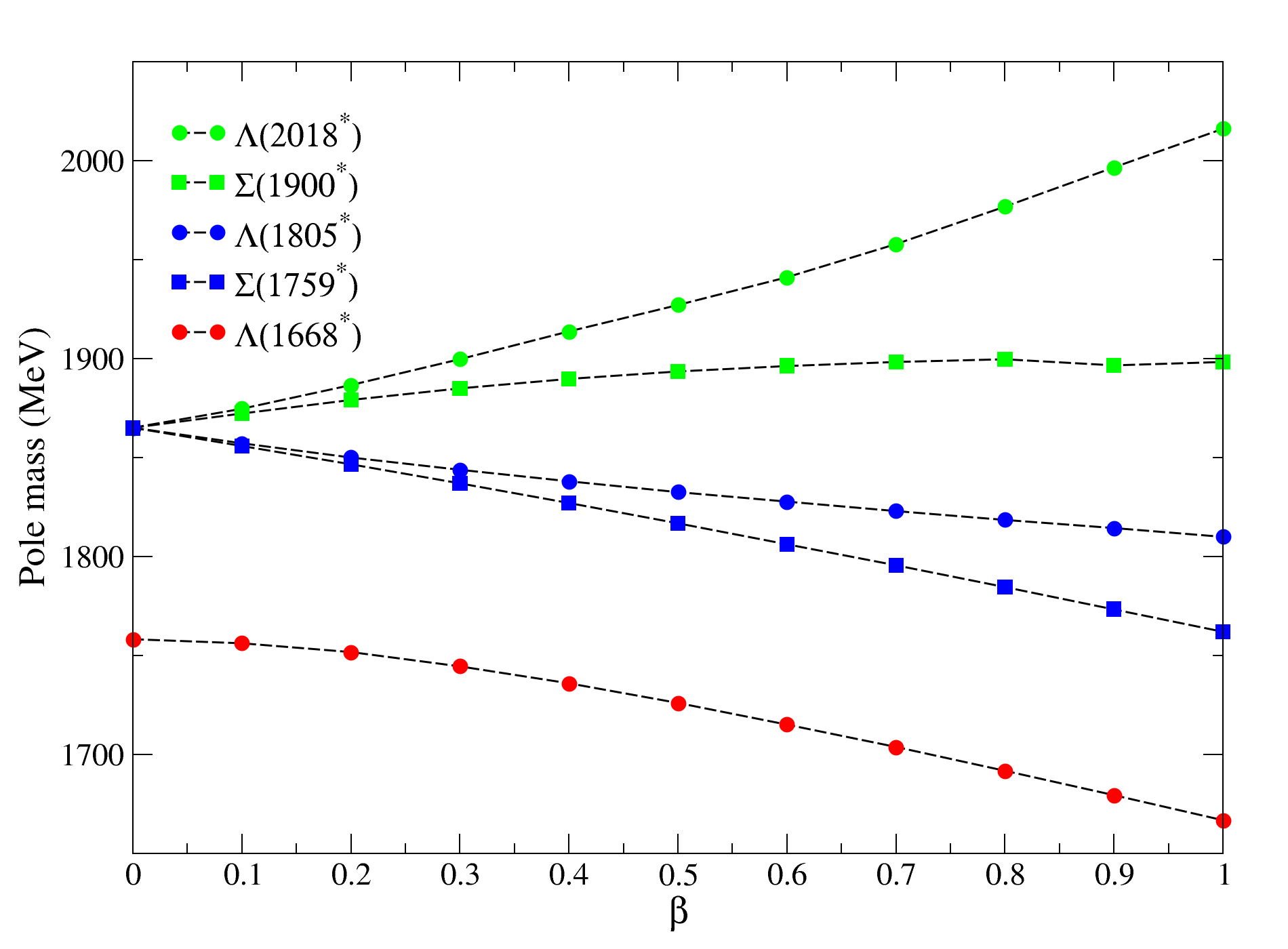}
    \caption{Real part of the pole position of the five found states in terms of the parameter $\beta$ controlling the SU(3) symmetry breaking (see Eq.~\eqref{eq:beta}). We have used  $\langle M \rangle = 1151.14$ MeV and $\langle m \rangle = 853.73$ MeV for the average masses of the baryon octet and meson nonet, respectively. For $\beta< 0.33$, all the states become bound states since they appear in the first Riemann sheet. }
    \label{fig:poles_SU3}
\end{figure}

In the SU(3) limit, states naturally organize into irreducible representations, making the identification of the singlet and two-octet states straightforward. In the present formalism, the SU(3) symmetry is only broken by the meson and baryon masses. To effectively restore SU(3) symmetry, we replace the physical masses of the hadrons ($M_i$ and $m_i$) involved in the relevant channels by their corresponding  (quasi-)symmetric counterparts ($\widetilde{M}_i$ and $\widetilde{m}_i$),
\begin{subequations}\label{eq:beta}\begin{align}
    M_i\ & \to\ \widetilde{M}_i = \langle M\rangle + \beta(M_i - \langle M\rangle) \hspace{0.4cm}\text{(for baryons)}\,,\\
    m_i\ & \to\ \widetilde{m}_i = \langle m\rangle + \beta(m_i - \langle m\rangle) \hspace{0.4cm}\text{(for mesons)}  \,.
\end{align}
\end{subequations}
Here, $\langle M \rangle$ and $\langle m \rangle$ represent the average masses of the baryon octet and meson nonet, respectively. The parameter $\beta \in [0,1]$ controls the degree of SU(3) symmetry breaking. Thus, $\beta = 0$ corresponds to the exact symmetric limit and for $\beta = 1$ the physical masses are recovered. In Fig.~\ref{fig:poles_SU3} the evolution of the masses of the five dynamically generated poles is shown as a function of $\beta$. The lowest-lying $\Lambda^*$ state corresponds to the singlet irrep, where the interaction is more attractive. Actually, in the SU(3) limit, its eigenvalue is twice larger than that of the two octet representations.

\begin{table*}
    \centering
    \begin{tabular}{c|c|cc|ccccc} \hline
         & \multicolumn{3}{c|}{LO HGS $VB$} & \multicolumn{5}{c}{RPP~\cite{ParticleDataGroup:2024cfk} and Belle~\cite{Belle:2022ywa}}  \\ \hline
         & Complex plane & \multicolumn{2}{c|}{Real axis} & \multirow{2}{*}{Name} & \multirow{2}{*}{$J^P$} & \multirow{2}{*}{Status} & \multirow{2}{*}{Mass [MeV]} & \multirow{2}{*}{Width [MeV]} \\ 
         & $M_R-i\Gamma_R/2$ [MeV]& $M_R$ [MeV] & $\Gamma_R$ [MeV] & & & & \\ \hline
         & & & & $\Sigma(1430)$ & $1/2^-$ & Belle  & $1435-1440$ & $10-30$ \\
         & & & & $\Sigma(1580)$ & $3/2^-$ & $*$   & $\approx 1580$ & $\approx 10$ \\
         & $-$ & $1759$ & $6.12$ & $\Sigma(1620)$ & $1/2^-$ & $*$ & $1600-1650$ & $40-100$ \\
         $I=1$ & & & & $\Sigma(1670)$ & $3/2^-$ & $****$ & $1665-1685$ & $40-100$ \\
         & $-$ & $1900$ & $49.44$ & $\Sigma(1750)$ & $1/2^-$ & $***$ & $1700-1800$ & $100-200$ \\
         & & & & $\Sigma(1900)$ & $1/2^-$ & $**$ & $1900-1950$ & $140-190$  \\ 
         & & & & $\Sigma(1910)$ & $3/2^-$ & $***$ & $1870-1950$ & $150-300$ \\\hline
         & & & & $\Lambda(1520)$ & $3/2^-$ & $****$ & $1519.42\pm 0.19$ & $15.73\pm 0.26$ \\ 
         & $1668$ & $1668$ & $0$ & $\Lambda(1670)$ & $1/2^-$ & $****$ & $1670-1678$ & $25-35$ \\
         & & & & $\Lambda(1690)$ & $3/2^-$ & $****$ & $1685-1695$ & $60-80$ \\
         $I=0$ & $1805 - 20.14\,i$ & $1805$ & $13.86$ & $\Lambda(1800)$ & $1/2^-$ & $***$ & $1750-1850$ & $150-250$  \\
         & & & & $\Lambda(2000)$ & $1/2^-$ & $*$ & $\approx2000$ & $\approx200$  \\ 
         & $2018 - 32.90\,i$ & $2017$ & $26.72$ & $\Lambda(2050)$ & $3/2^-$ & $*$ & $\approx2050$ & $\approx500$ \\ [1mm]\hline
    \end{tabular}
    \caption{Properties of the five generated $\Lambda^*$ and $\Sigma^*$ resonances obtained in this work and their possible RPP counterparts. We have also considered the $\Sigma(1430)$ from a clear signal reported recently by the Belle collaboration in the $\Lambda\pi^\pm$ invariant  mass distributions measured in the  $\Lambda_c^+ \to \Lambda \pi^+\pi^+\pi^-$ decay.   However, the lowest-energy  $J^P=1/2^-$ $\Lambda(1380)$ and $\Lambda(1405)$ resonances are not included in the table,  since they form a well established double-pole structure generated by the $\pi\Sigma,\bar K N,\eta \Lambda$ and $K\Xi$ coupled-channel chiral interactions.  We recall that the UV cutoff $\Lambda=830$ MeV has been tuned to optimize the description of the $\Sigma^*$ resonances.}
    \label{tab:resonances}
\end{table*}

In summary, based on a purely group-theoretical analysis, one expects a strongly bound $\Lambda^*$ state corresponding to the singlet, along with two $\Lambda^*$ and two $\Sigma^*$ states arising from the two attractive octets. In fact, this expectation is confirmed in Fig.~\ref{fig:isospin_amplitude}. In addition, one has to bear in mind the accidental spin-parity $(1/2^--3/2^-)$ degeneracy that appears in the LO HGS formalism, which would give poles at the same location in both sectors.

\subsubsection{Spectrum}
\paragraph{Experimental information.}
  In Table~\ref{tab:resonances}, we  compile the positions of the poles obtained in this work, and their possible associations with the $J^P=(1/2^-,3/2^-)$ $\Lambda^*$ and $\Sigma^*$ states listed in the RPP~\cite{ParticleDataGroup:2024cfk}. In the isovector sector, we observe in the experimental spectra one  $(1/2^- - 3/2^-)$  pair of quasi-degenerate states, given the current precision, around 1900 MeV. In the region of 1700 MeV, there appear also two $(1/2^- - 3/2^-)$ states, but in this case the spin-breaking shift is larger, about $85\pm 50$ MeV. Furthermore, the $3/2^-$ resonance appears to be lighter than the $1/2^-$ spin one, just the opposite of what is inferred from the candidates that would form the lightest doublet. 
  
  The one-star $\Sigma(1580)$ and $\Sigma(1620)$ resonances are also listed in the RPP, with spin-parities $3/2^-$ and $1/2^-$, respectively. In the RPP classification of baryons, a one-star indicates that the particle has been observed and its properties have been measured, but its mass or spin has not yet been determined with sufficient precision to give it a two-star or three-star classification. Some evidences for these two resonances were reported in the 70's, however these states have not been seen in any of the most recent experimental analyses.  Actually, the BNL experiment of Ref.~\cite{Carroll:1976gs} reported in 1976 two bumps in the $I=1$ $K^- N$ total cross section in this energy region. However, the signals have not been confirmed in recent experiments~\cite{ParticleDataGroup:2024cfk}. On the other hand, the Belle Collaboration has very recently reported the first observation of $\Lambda\pi^+$ and $\Lambda \pi^-$ signals near the $\bar K N$ mass threshold, around 1435-1440 MeV, in the $\Lambda_c^+ \to \Lambda \pi^+\pi^+\pi^-$ decay. The significances for the observed signals were large, 7.5$\sigma$ for $\Lambda\pi^+$ combination, and $6.2\sigma$ for $\Lambda\pi^-$ combination, respectively. However, limited by the statistics and the shape of the background, the Belle Collaboration could not distinguish between $J^P=1/2^-$ $\Sigma(1430)$ resonances with a width about 10-30 MeV and $\bar K N$ threshold cusps, since both fits give similar $\chi^2$.

  In the 1.7-2.0 GeV region of energies, the  RPP odd-parity  $\Lambda^*$ spectrum contains three $(1/2^- - 3/2^-)$ doublets, though the higher one located around 2 GeV would be formed by two resonances whose  evidence of existence is poor (one-star). In addition, there are the two $1/2^-$ $\Lambda(1380)$ and $\Lambda(1405)$ resonances that form a well-established double-pole structure originating from the  $\pi\Sigma,\bar K N,\eta \Lambda$ and $K\Xi$ coupled-channel chiral interactions. We do not include these two states in Table~\ref{tab:resonances}, since they are far below the $VB$ thresholds in the $S=-1$ sector, and the $VB$ degrees of freedom should play a minor role in the dynamics of this double pole.

\paragraph{Other hadron-molecular degrees of freedom.}

In addition to $VB$ scattering, $J^P=1/2^-$ and $3/2^-$ $S-$wave resonances can be generated by the  chiral interaction  between Goldstone-bosons of the pion-octet and the $1/2^+$ and $3/2^+$ baryons of the nucleon-octet ($PB$) and $\Delta-$decuplet ($PB_{3/2}$). At LO,  they are given by the WT Lagrangian which provides coupled-channel potentials, projected onto isospin-strangeness sectors,  of the type of Eq.~\eqref{eq:new_kernel} for the $VB$ interaction, 
\begin{equation}
    V^{P{\cal B}}_{ij} = -\frac{C^{P{\cal B}}_{ij}}{4f^2}  \sqrt{\frac{M_i+E_i}{2M_i}} \sqrt{\frac{M_j+E_j}{2M_j}} (2\sqrt{s} - M_i - M_j) 
    \label{eq:WT}
\end{equation}
with ${\cal B}=B,B_{3/2}$. In the $PB$ case, the SU(3) WT  $C^{PB}$  matrix has three attractive eigenvalues, corresponding to two octets and  one singlet irreps~\cite{Oller:2000fj, Garcia-Recio:2003ejq}, which for the $\Lambda^\ast$ sector  give rise to the double-pole structure of the $\Lambda(1405)$~\cite{ParticleDataGroup:2024cfk} and they should also play a significant role in the dynamics of the $\Lambda(1670)$~\cite{Garcia-Recio:2002yxy,Jido:2003cb, Garcia-Recio:2003ejq,Nieves:2024dcz}.  On the other hand, the  $PB$ ($\pi \Lambda, \pi\Sigma,\bar K N,\eta \Sigma,K\Xi $) WT $S-$wave coupled-channel dynamics also produce visible signatures in the isovector $J^P=1/2^-$ amplitudes, as expected from the attractive octet irreps~\cite{Oller:2000fj,Jido:2003cb,Garcia-Recio:2003ejq}, which should be related to the $\Sigma(1430)$ reported by Belle, and the one-star $1/2^-$ $\Sigma(1620)$ resonance compiled in the RPP.  

In the $PB_{3/2}$ sector, the SU(3) reduction of the WT matrix reads: $8 \otimes 10 = 35\oplus 27 \oplus 10 \oplus 8$. The interaction is attractive in the octet, decuplet and 27-plet irreps but repulsive in the 35-plet channel~\cite{Kolomeitsev:2003kt,Sarkar:2004jh}, with eigenvalues 6,3,1 and $-3$, respectively. To simplify the discussion we will not consider the very weakly attractive 27-plet.\footnote{Nevertheless, this irrep might have some influence on the $\Lambda(1690)$ dynamics, as claimed in~\cite{Kolomeitsev:2003kt}. Moreover, constituent quark model (CQM) degrees of freedom could be also relevant for this resonance~\cite{Nieves:2024dcz,Yoshida:2015tia}.} Thus, one might generate   $3/2^-$ $\Sigma^*$ resonances from the octet and decuplet interactions, and a $1/2^-$ $\Lambda^*$ from the octet. The isoscalar one is expected to be a component  of the physical  $\Lambda(1520)$ \cite{Kolomeitsev:2003kt,Sarkar:2004jh}, while the isovector states might be related to to the five-star $\Sigma(1670)$ and $\Sigma(1910)$ resonances \cite{Kolomeitsev:2003kt,Sarkar:2004jh} or even to the one-star $3/2^-$ $\Sigma(1580)$ that, as in the case of  the  $1/2^-$ $\Sigma(1620)$, was reported in the 70's, but it has not been seen in subsequent analyses~\cite{ParticleDataGroup:2024cfk}. 

The $VB_{3/2}$ pairs in $S-$wave can also couple to $1/2^-$ and $3/2^-$, but all $(S=-1)-$thresholds are above 2.1 GeV. We will mostly ignore these channels in the discussion, though some of them might play some role in the dynamics of the  higher energy states collected in Table~\ref{tab:resonances}, as claimed in Refs.~\cite{Gamermann:2011mq,Sarkar:2010saz}.  

In the HGS formalism, the mixing between $PB$, $PB_{3/2}$, $VB$ and $VB_{3/2}$ channels is commonly neglected, and each meson-baryon sector is separately analyzed. This is justified in the $P{\cal B}-V{\cal B}$ case since,  within the HGS scheme, such mixing is driven by the $VVP$ vertex which is suppressed (see the discussion in Ref.~\cite{Dias:2021upl}).  The SU(6) spin-flavor approach of Ref.~\cite{Gamermann:2011mq} considers the coupling among  all possible combinations of pseudoscalar and vector mesons with the ground-state octet and decuplet baryons.  The interaction of the Goldstone bosons and baryons is identical in both schemes, since it is determined by chiral symmetry. The differences arise when vector mesons are involved because, in addition to the $0^+$ $t-$exchange,\footnote{It corresponds to the zero component of the exchanged  vector meson, and  it leads to a contact term when the three-momenta of the external hadrons are neglected compared to the typical mass $M_V$ of the  vector meson.} the axial-vector $1^+$ exchange is also present in Ref.\,\cite{Gamermann:2011mq}, as required by SU(6) symmetry, while it is not present in the HGS formalism. 

\paragraph{Discussion.} The identification of the dynamically generated resonances found in this work with some of the RPP ones compiled here  in Table~\ref{tab:resonances} is not straightforward. Physical states should be result of the interplay between different  hadron-molecular degrees of freedom and/or possible CQM components.\footnote{See for instance the analysis in Ref.~\cite{Nieves:2024dcz} for the odd-parity $\Lambda_{Q=b,c,s}$ states or the study carried out in Ref.~\cite{Cincioglu:2016fkm} for the exotic $X(3872)$ and its heavy-quark spin-flavor partners.}  Within unitarized approaches, the effects of degrees of freedom not explicitly considered are expected to be approximately accounted for  the finite parts of the renormalized two-hadron loops. The latter are controlled by the UV cutoff, set in this study at the reasonable value of 830 MeV (see discussion of Refs.~\cite{Nieves:2024dcz,Guo:2016nhb,Albaladejo:2016eps}). In addition, irreducible NLO contributions~\cite{Feijoo:2018den,Feijoo:2023wua,Nieves:1999bx} to the BSE kernel potential also account for effects from these other degrees of freedom.

 We start the discussion in the $\Sigma^*$ sector. It seems quite reasonable to associate the degenerate $(1/2^--3/2^-)$ pole found in this work at $1759$~MeV with the $\Sigma(1670)$ and $\Sigma(1750)$ resonances, and clearly the one at $1900$~MeV with the approximate $\Sigma(1900)$ and $\Sigma(1910)$ spin-doublet. The latter assignment is an important success of the LO HGS approach to $VB$ interactions in this sector, and supports the choice of 830 MeV for the UV cutoff.  

Within this interpretation, the $1/2^-$ lower-energy one-star $\Sigma(1620)$, if it exists, and the new $\Sigma(1430)$ would be mostly generated by the two attractive octets of the WT $PB$ coupled-channel dynamics. This is well known since the first results from the first UChPT analyses carried out in  Refs.~\cite{Oller:2000fj,Jido:2003cb, Garcia-Recio:2003ejq}, where it was already  shown how the intricate effects of coupled channel dynamics and of the SU(3) mass breakdown in the isovector sector produce broad poles in the unitarized WT $PB$ amplitudes, close to the antikaon-nucleon threshold, but not located in the appropriate non-physical Riemann sheets. These poles lead  to visible peaks in the physics amplitudes, though they might not qualify as  proper resonances.\footnote{As already mentioned, the width of the $\Sigma(1430)$ state is about $10–30$ MeV according to the Belle experiment \cite{Belle:2022ywa}, which at first sight seems difficult to reconcile with the imaginary parts ($\gtrsim$ 100 MeV) found in the UChPT approaches. However, poles found around thresholds must be taken with caution, and one should note that the experiment will see amplitudes in the physical real energy axis and in that case the extrapolation to the complex plane will be very model dependent. This is actually the case with the $\Sigma(1430)$ as discussed in Ref.~\cite{Li:2024tvo}. There it is shown that in spite of having a pole indicating a width of about 200 MeV, the results of $|T_{ij}|^2$ just produce a peak around the $\bar K N$ threshold with an apparent width of 30–50 MeV, as seen in the Belle experiment. Within this context, the nature of the $\Sigma(1430)$ state is investigated in Ref.\,\cite{Lin:2025pyk}.} 
       
On the other hand, the WT $PB_{3/2}$ coupled-channel chiral interaction could produce two dynamically generated $3/2^-$ resonances stemming from the attractive octet and decuplet irreps. They may be related to the one-star $\Sigma(1580)$ and have also influence on the physical five-star $\Sigma(1670)$. Nor can it be ruled out that there is another state that has not yet been discovered.

All these chiral $P{\cal B}$ channels could interfere with the $VB$ ones considered here and  should explain  the spin-breaking between the 
$\Sigma(1670)$ and $\Sigma(1750)$ physical states.  

 Now we move to the $\Lambda^*$ sector, where we found three poles.  The sharp UV cutoff  was tuned to the isovector spectrum, and hence the isoscalar states become a prediction. The degeneracy encoded in the $I=0$ poles located at $1668$~MeV and at $2018$~MeV could correspond to the $\Lambda(1670)-\Lambda(1690)$ and the $\Lambda(2000)-\Lambda(2050)$, respectively. With this assumption, one concludes that the pole appearing at $1805$~MeV should be associated to the $\Lambda(1800)$ and to its  $3/2^-$ sibling not discovered yet. This conjecture is supported by the closeness of the theoretical poles to the experimental masses, although it seems little compatible with  a large (dominant) chiral $PB$ components in the $\Lambda(1670)$ wave-function. Indeed the  largest decay modes of this resonance are $\bar K N$, $\pi \Sigma$ and $\eta \Lambda$ with $(20-30)\%, (25-55)\%$ and (10-25)\% fractions, respectively. One would expect that such non-$VB$ contributions should produce a much larger  $(1/2^--3/2^-)$ mass breaking.

\begin{figure*}[t]
\includegraphics[width=15cm,keepaspectratio]{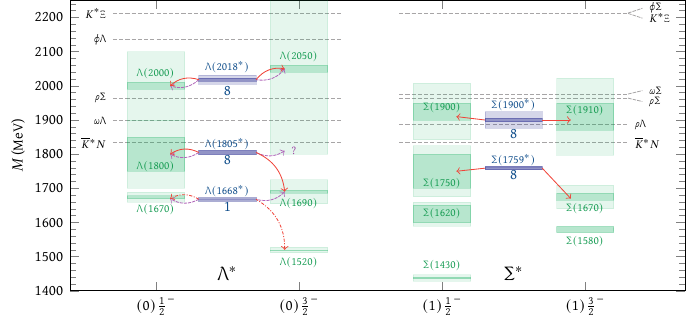}
\caption{Schematic representation of the odd-parity $\Lambda^*$ and $\Sigma^*$ experimental states in Table\,\ref{tab:resonances} and their assignment to the poles predicted in this work. The green boxes represent the experimental masses. The inner, darker boxes represent the uncertainty in the mass of the resonances, while the outer, lighter ones span their experimental half-width (in both directions). The blue boxes are the theoretical determinations in our work. The purple (red) arrows represent the first (second) assignment discussed in the text, to which the reader is referred for further details. \label{fig:scheme-assign}}
\end{figure*}

Admitting a larger $(1/2^--3/2^-)$ mass-splitting allows for alternative interpretation. Based on Fig.~\ref{fig:poles_SU3}, one can draw an analogy between the $\Sigma(1759*)$ and the $\Lambda(1805^*)$ poles, as well as between the $\Sigma(1900^*)$ and the $\Lambda(2018^*)$ ones, suggesting that each of these pairs stem from different SU(3) octet irreps. This, combined with the ($3/2^-$–$1/2^-$, $1/2^-$–$3/2^-$) pattern observed in the $\Sigma^*$ resonances listed in Table~\ref{tab:resonances}, supports the association  of the poles at 1805 MeV and 2018 MeV  to the $\Lambda(1690)$--$\Lambda(1800)$ and the $\Lambda(2000)$--$\Lambda(2050)$ pairs, respectively. Large CQM~\cite{Yoshida:2015tia, Nieves:2024dcz} and/or SU(3) 27-plet $PB_{3/2}$ components in the $3/2^-$ $\Lambda(1690)$ dynamics might be responsible for the sizable spin-breaking between the  $\Lambda(1690)$ and $\Lambda(1800)$ physical states. Additionally  in this picture,  the $\Lambda(1668^*)$ pole, which   corresponds to the SU(3) singlet irrep, would provide components to the $1/2^-$ $\Lambda(1670)$ and $3/2^-$ $\Lambda(1520)$  physical states. For instance, it is well established the important role played by the coupling of the $\Lambda(1520)$ to the $\bar K^* N$ pair, the largest one for the the $\Lambda(1668^*)$ pole, in the photo-production of this resonance~\cite{Toki:2007ab}. We note however  that both the $\Lambda(1670)$ and  $\Lambda(1520)$ physical states should contain very relevant contributions from the chiral $PB$ and $PB_{3/2}$ channels, respectively, as commonly accepted in the literature. Quark models also predict states in the vicinity of these two resonances~\cite{Yoshida:2015tia}.

The two proposed assignments are schematically depicted in Fig.\,\ref{fig:scheme-assign}. Both appear plausible as interpretations of the $I=0$ poles in terms of experimental $\Lambda^*$ states and warrant further consideration. We believe, however, that the second assignment, where the two higher-energy isoscalar poles are identified with the $\Lambda(1690)$--$\Lambda(1800)$ and $\Lambda(2000)$--$\Lambda(2050)$ pairs, is likely more reliable.

To conclude the comparison between the theoretical states and the experimental $\Lambda^*$ and $\Sigma^*$ resonances, it should be mentioned that, although the theoretical model does not fully reproduce the experimental widths, likely due to the omission of certain decay channels, the resulting widths remain in reasonable agreement with the order of magnitude of those reported in the RPP.

\subsection{Scattering parameters}

For completeness, we compile in Table~\ref{tab:scatteringparam} the values found for the scattering parameters, introduced in Subsec.~\ref{sec:sp}, and computed including and omitting the widths of the $\rho$ and $\bar{K}^*$ mesons. The most significant impact of including the vector meson widths is observed in the scattering length $a_{0,i}$ and effective range $r_{0,i}$ of the $\rho^+\Lambda$ channel. This effect can be attributed to the inclusion of the $\rho$ meson width, which shifts the position of the $\Sigma(1900)^*$ pole, making it 2\,MeV more bound. As a result, when such a width is omitted in the calculations, the $\rho^+\Lambda$ elastic amplitude experiences an effective enhancement near its corresponding threshold. These values may serve as a useful reference for future experimental analyses. We only show the scattering parameters of the channels for which we compute the CFs in the next subsection.

\begin{table}[t]
    \centering
    \begin{tabular}{ccc} \hline
        \multirow{2}{*}{Channel} & \multicolumn{2}{c}{$a_0$ [fm]} \\
         & With Widths & Without Widths \\ \hline
        $\bar{K}^{*0}p$       & $\hphantom{+}0.711  -0.227\,i$ & $0.616$ \\
        $\rho^+\Lambda$       & $-0.078-0.272\,i$  & $-0.277-0.005\,i$ \\
        $\rho^0\Sigma^+$      & $\hphantom{+}0.591-0.395\,i$   & $\hphantom{+}0.511-0.021\,i$ \\
        \hline
        $\omega\Lambda$       & $\hphantom{+}0.150-0.030\,i$   & $\hphantom{+}0.147-0.029\,i$ \\
        $\rho^0\Sigma^0$      & $\hphantom{+}0.468 -0.156\,i$ & $\hphantom{+}0.374 -0.004\,i$ \\
        $\phi\Lambda$         & $\hphantom{+}0.228 -0.086\,i$ & $\hphantom{+}0.229 -0.085\,i$ \\ \hline
    \end{tabular} \vspace{0.2cm}
    
    \begin{tabular}{ccc} \hline
        \multirow{2}{*}{Channel} & \multicolumn{2}{c}{$r_0$ [fm]} \\
         & With Widths & Without Widths \\ \hline
        $\bar{K}^{*0}p$       & $-1.577 - 0.138\,i$   & $-0.495 $ \\
        $\rho^+\Lambda$       & $-6.201 + 12.76\,i$     & $-41.31-1.460\,i$ \\
        $\rho^0\Sigma^+$      & $-1.201 + 0.127\,i$   & $-1.049 + 0.122\, i$ \\
        \hline
        $\omega\Lambda$       & $-2.703 - 1.303\,i$ & $-1.717 -1.998\,i$ \\
        $\rho^0\Sigma^0$      & $-0.715 - 0.015\,i$   & $-0.219 -0.183\,i$ \\
        $\phi\Lambda$         & $-0.200 - 0.367\,i$ & $-0.187 - 0.342\,i$ \\ \hline
    \end{tabular}
    \caption{Scattering parameters $a_{0,i}$ (top) and $r_{0,i}$ (bottom), defined in Eq.~\eqref{eq:a0r0}, for the six channels whose CFs will be discussed in Subsec.~\ref{sec:results_CF}. The parameters have calculated both including and neglecting the effects due to the consideration of the $\rho$ and $\bar{K}^*$ meson widths. }
    \label{tab:scatteringparam}
\end{table}

\subsection{Correlation functions}
\label{sec:results_CF}

We now present, for the first time, predictions for various $VB$ CFs in the $S=-1$ sector. These are analyzed in the physical basis of the relevant channels, with particular attention given to those that exhibit significant proximity or coupling to the five dynamically generated states in the $I = 0$ and $I = 1$ sectors. In this way, future comparisons with corresponding experimental results may offer novel insights into the underlying theoretical spectroscopy. These  are the $\bar{K}^{*0}p$, $\rho^+\Lambda$ and $\rho^0\Sigma^+$ channels with electric charge $Q=+1$, and the neutral ($Q=0$) $\omega\Lambda$, $\rho^0\Sigma^0$ and $\phi\Lambda$ ones. We do not explicitly study charged channels in $Q=0$ to avoid introducing Coulomb interactions, but they are included in the coupled-channel dynamics where the electromagnetic interaction is negligible. Note that working in physical basis involves a change of basis of the $C_{ij}$ coefficients entering in Eq.~\eqref{eq:interactionkernel}. The coefficients in the physical basis for the $Q = +1$ and $Q = 0$ channels are shown in Appendix \ref{app:coefficients} in the left and right panels of Table~\ref{tab:Q coefficients}, respectively. We use the realistic production-weights compiled in Table~\ref{tab:weights}.

\begin{figure*}
    \centering
    \includegraphics[width=0.95\textwidth]{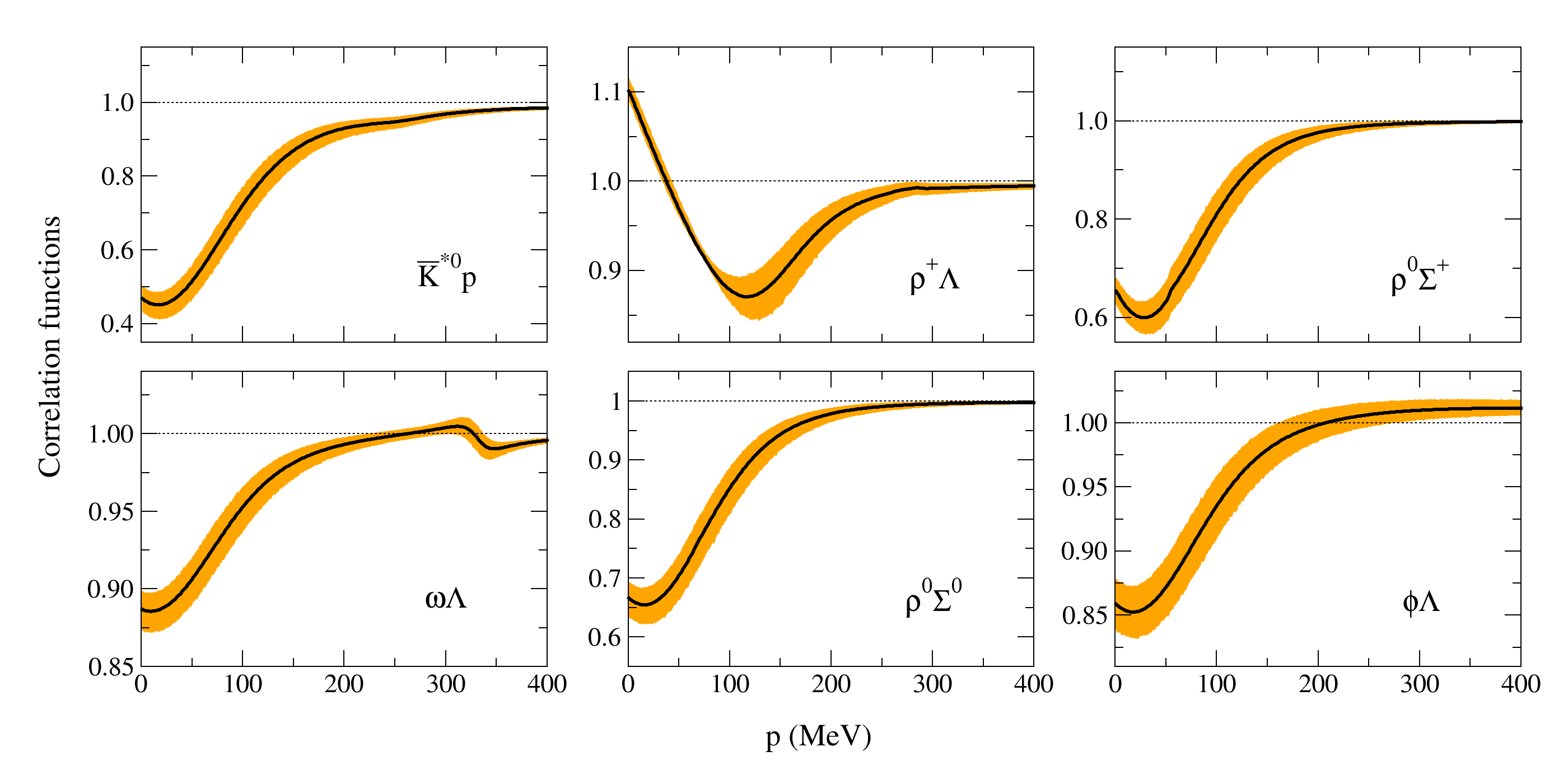}
    \caption{CFs for different physical hadron pairs. Predictions for the $Q=+1$ ($Q=0$) $\bar{K}^{*0}p$, $\rho^+\Lambda$ and $\rho^0\Sigma^+$ ($\omega\Lambda$, $\rho^0\Sigma^0$ and $\phi\Lambda$) channels are displayed in the top (bottom) panels. The orange 68\% CL bands  are estimated from the production-weights and source-radii uncertainties.}
    \label{fig:CFs}
\end{figure*}

The calculated CFs are displayed in Fig.~\ref{fig:CFs}. The 68\% confidence-level (CL) bands are estimated using a Monte Carlo sampling approach to propagate  the uncertainties from the production weights (Table~\ref{tab:weights}) and the source radii (see Section~\ref{sec:correlationfunction}). In these plots, the presence of resonances below thresholds is reflected in the general shape of some CFs at low momenta, similarly to those seen in previous works (see Refs.\,\cite{Liu:2023uly,Sarti:2023wlg,Albaladejo:2023pzq,Encarnacion:2025lyf}).  

In the upper panels of Fig.~\ref{fig:CFs}, the impact of the $I=1$ resonances is expected to be clearly visible, given the isospin content of the  considered channels. The $\Sigma(1759^\ast)$ state lies about $70\,\text{MeV}$ below the threshold of the $\bar{K}^{\ast0}p$ channel, to which it strongly couples, making the CF of this pair quite sensitive to the presence of this state. The $\rho^+\Lambda$ and $\rho^+\Sigma^0$   thresholds are located too far above the $\Sigma(1759^*)$ and their CFs do not show significant signatures related to this state.  The $\Sigma(1900^*)$, on the contrary, is visible in all three upper CFs of Fig.~\ref{fig:CFs}. In the $\rho^0\Sigma^+$ one, this state appears approximately 60~MeV below threshold and influences the CF behavior  at low momenta, as clearly seen in the plot.  The \(\rho^+\Lambda\) channel is particularly interesting. The attractive nature of the interaction near the threshold, as evidenced by its scattering length in Table~\ref{tab:scatteringparam}, is reflected in a CF that exceeds unity at low momenta. As the momentum increases, the CF gradually drops below one, signaling the onset of repulsive features. In particular, a noticeable depletion around $p \sim 120\,\text{MeV}$ can be attributed to the presence of the $\Sigma(1900^*)$ resonance, whose influence becomes apparent in this momentum region. Similarly, the $\Sigma(1900^*)$ state leaves only a faint trace near $p\sim 250$~MeV in the $\bar{K}^{*0}p$ CF. Furthermore, in the upper panels of Fig.~\ref{fig:CFs}, two tiny cusps are visible in the $\rho^+\Lambda$ and $\rho^0\Sigma^+$ CFs at $p \sim 280$ and $50$~MeV, respectively. These features correspond to the opening of the $\omega\Sigma$ and $\rho^+\Sigma^0$ channels.

\begin{figure*}
    \centering
    \includegraphics[width=0.95\textwidth]{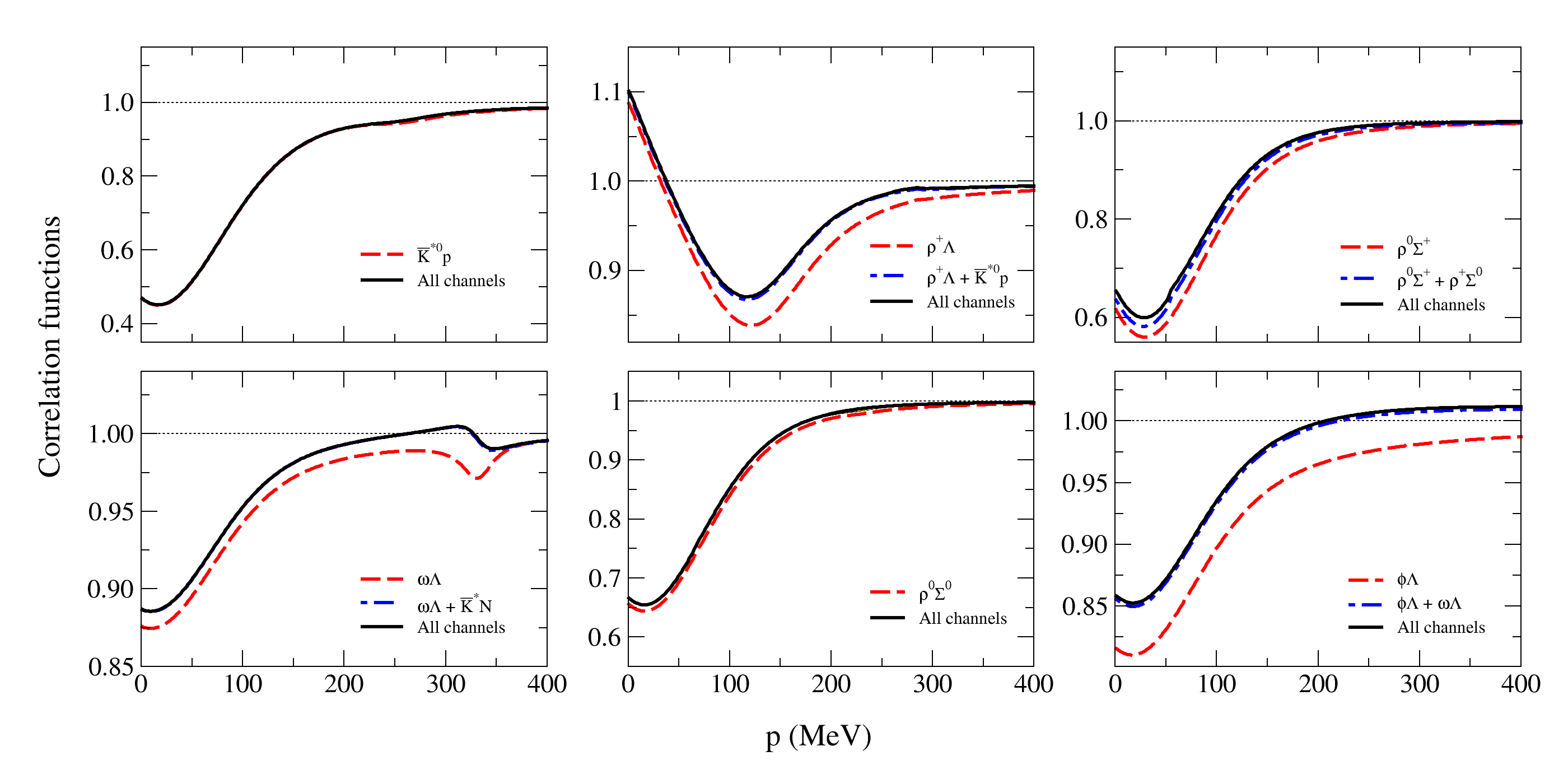}
    \caption{Different coupled-channel contributions to the CFs displayed in Fig.~\ref{fig:CFs}.}
    \label{fig:CFs_CHAN}
\end{figure*}

Next we discuss the CFs displayed in the bottom panels of Fig.~\ref{fig:CFs}, which are mostly sensitive to the isoscalar part of the $VB$ interaction. The lightest channel is $\omega\Lambda$, which  threshold is located approximately 230~MeV above the dynamically generated $\Lambda(1668^*)$ state. Owing to this substantial binding energy, no discernible signal from this state is expected to manifest in any of the three CFs displayed. On the other hand, the $\Lambda(1805^*)$ pole lies below but close enough to the $\omega\Lambda$ and $\rho\Sigma$ thresholds, and may therefore leave a visible imprint on their CFs. However, this is difficult to confirm, as the shape of these CFs could either be a consequence of the pole’s presence or the result of a repulsive interaction near threshold, as can be inferred from the respective scattering lengths listed in Table~\ref{tab:scatteringparam}. An analogous situation is observed for the $\Lambda(2018^*)$ pole with respect to the $\phi\Lambda$ CF. In contrast, the $\Lambda(2018^*)$ state produces a clear signature in the $\omega\Lambda$ CF, visible as a structure around $p\sim 300$~MeV, indicative of a strong coupling to this channel. However, no such effects are observed in the $\rho^0\Sigma^0$ CF, in agreement with the small coupling of the $\Lambda(2018^*)$ to this channel, as shown in Table~\ref{tab:couplings_I0}. Taken together, these features highlight the sensitivity of the CFs to the interplay between resonance properties (mass, width, and coupling strength) and the nearby hadron-hadron thresholds.

Still referring to the lower panels of Fig.~\ref{fig:CFs}, a common feature of the three selected channels is  that the corresponding elastic interaction kernel $C_{ii}$ (Table~\ref{tab:Q coefficients}) is zero. This implies that the dynamics is entirely governed by coupled-channel effects. The coupled-channel dynamics is subsequently mapped onto the CFs via the KP formula (Eqs.~\eqref{eq:kooninpratt} and \eqref{eq:wf_ampli}).  Consistently, the resulting $\omega\Lambda$, $\rho^0\Sigma^0$, and $\phi\Lambda$  CFs exhibit moderate deviations from unity, as evidenced by the shallow dips observed in each case. The similarity in the shape and magnitude of these deviations, together with the absence of clear  signatures from the $\Lambda(1805^*)$ and $\Lambda(2018^*)$ states near threshold, further supports the interpretation of a predominantly repulsive elastic interaction in these channels.

 The contributions of the different coupled channels entering in Eq.~\eqref{eq:kooninpratt} through $\psi_{ji}$, each one scaled by its production weight, are displayed in Fig.~\ref{fig:CFs_CHAN}.\footnote{Even if $V_{ii}$ were zero, the elastic wave function $\psi_{ii}$ is not zero and is fully determined by the coupled channel effects that determine $T_{ii}$.} In general, it can be seen that the inelastic contributions to these CFs are very small, being at most a 7\% of the total CF in the case of the $\rho^0\Sigma^+$ and even smaller for the other ones. For the $\bar{K}^{*0}p$, the elastic interaction kernel is non-zero (see Table~\ref{tab:Q coefficients}), and  the diagonal  contribution of the KP CF greatly dominates over the non-diagonal ones. For the rest of CFs, the  elastic potential coefficient is zero and their elastic transition relies on the rescattering. In these cases, although the elastic transition is dominant, one can see that the most important inelastic contributions correspond precisely to the channels with a non-zero $C_{ij}$ coefficient (see Table~\ref{tab:Q coefficients}). This is the case of the $\bar{K}^{*0}p \to\rho^+\Lambda$ contribution in the $\rho^+\Lambda$ CF, the $\rho^0\Sigma^+ \to \rho^+\Sigma^0$  or the $\omega\Lambda\to \bar{K}^*N$ ones in the $\rho^0\Sigma^+$ and   $\omega\Lambda$ CFs, respectively. The case of the $\phi\Lambda$ CF is trickier, because the only inelastic transition that seems to play an important role in the total CF is the $\omega\Lambda\to\phi\Lambda$ one, even though the coefficient of the associated  potential ($C_{ij}$) is zero.  The only non-zero couplings of the $\phi\Lambda$ pair are to the $\bar{K}^*N$ and $K^*\Xi$ channels. The $K^*\Xi$ production weights are negligible (see Tab.~\ref{tab:weights}), while the $\bar{K}^*N\to\phi\Lambda$ contribution to the $C_{\phi\Lambda}$ is largely suppressed by the distance between their respective thresholds. The $\omega\Lambda$ contribution in this CF, coming entirely from rescattering processes, turns out to be the most relevant inelastic one. It is also worth mentioning the large production weight associated with the inelastic transition $\omega\Lambda \to \phi\Lambda$ ($w_j=4.56$, see the last column of Table~\ref{tab:weights}), which further enhances the strength of this contribution.

\section{Conclusions}
\label{sec:conclusions}
The study of QCD at low energies requires the combined use of alternative techniques to the usual perturbative methods, such as LQCD simulations, unitarized/dispersive coupled-channel EFTs  and/or phenomenological analyses. In this context, the QCD spectrum contains not only the states described by the CQM, but also possible exotic compact multi-quark states (tetraquarks, pentaquarks) or hadronic molecules, which are extremely sensitive to the low energy parameters of the scattering amplitudes, and hence to the LECs of the EFTs.  Due to the short lifetime of many unstable hadrons, it is not possible to study their interactions by traditional scattering experiments, which greatly complicates the study of the hadron spectrum. As an alternative, LQCD and weak decays of heavy hadrons have been key tools to study new hadronic states, although they present limitations due to the complexity of the decay mechanisms and/or the high computational cost of the required calculations.  Hadron femtoscopy   overcomes some of these limitations, and the measured CFs contain  extremely  valuable information on  hadron-hadron interactions. Therefore, the formulation of a systematic improvable treatment of off-shell ambiguities of the scattering quantum wave-function within the KP formalism is a major theoretical challenge at this time.

In this work, we have revisited  the $S=-1$ sector of the interaction between vector mesons and ground state baryons, within the unitary coupled-channel HGS formalism. We have employed a scale-free dimensional regularization, which induces a reasonable short-distance behavior of the two-hadron wave functions~\cite{Molina:2025lzw}.  We have performed an exhaustive spectroscopy study, implementing different improvements and considerations, and  have found compatibilities between the poles extracted from the present revisited approach, and some of the states listed in the RPP compilation.  Finally, we have predicted the CFs associated with various meson-baryon pairs in this sector, which could be measured by ALICE in the near future, paying special attention to the distinctive and clear signatures produced by dynamically generated states. To obtain CF estimates as realistic as possible,  we have also calculated the production weights, which largely differ from one, for high-multiplicity events employing the production yields obtained using the Thermal-FIST package \cite{Vovchenko:2019pjl} and relative momentum distributions extracted within the CBW model \cite{Schnedermann:1993ws}.

The direct comparison of these theoretical predictions with data will provide further insights and constrains of the $VB$ interactions, including the possible interplay with CQM degrees of freedom and the mixing with $VB_{3/2}$, $PB$ and $PB_{3/2}$ sectors, which involve Goldstone bosons and baryons of the $\Delta-$decuplet.  Such studies will shed light into the odd-parity $\Lambda^*$ and $\Sigma^*$ hadron spectra, up to 2 GeV, and will open the strategy to improve EFTs by using LECs fitted both to LQCD calculations and femtoscopy data.

\begin{acknowledgements}
This work is part of the Grants PID2020-112777GB-I00, PID2023-147458NB-C21 and CEX2023-001292-S of MICIU/AEI /10.13039/501100011033 and of the Grant  CIPROM 2023/59 of Generalitat Valenciana. %
M.\,A. acknowledges financial support through GenT program by Generalitat Valencia (GVA) Grant No.\,CIDEGENT 2020/002, Ramón y Cajal program by MICINN Grant No.\,RYC2022-038524-I, and Atracción de Talento program by CSIC PIE 20245AT019. %
M.\,A.\, and A.\,F.\, thank the warm support of the ACVJLI.
\end{acknowledgements}

\appendix

\section{\boldmath $C_{ij}$ coefficients}
\label{app:coefficients}

\begin{table*}
    \resizebox{0.95\textwidth}{!}{
    \centering
    \begin{tabular}{cccccccc} \hline
        $Q=+1$ & $\bar{K}^{*0} p$ & $\rho^+\Lambda$ & $\rho^0\Sigma^+$ & $\rho^+\Sigma^0$ & $\omega\Sigma^+$ & $\phi\Sigma^+$ & $K^{*+}\Xi^0$  \\ \hline
        $\bar{K}^{*0} p$ & $1$ & $\sqrt{3/2}$ & $-\sqrt{1/2}$ & $\sqrt{1/2}$ & $\sqrt{1/2}$ & $-1$ & $0$  \\
        $\rho^+\Lambda$ &  & $0$ & $0$ & $0$ & $0$ & $0$ & $\sqrt{3/2}$  \\
        $\rho^0\Sigma^+$ & & & $0$ & $-2$ & $0$ & $0$ & $\sqrt{1/2}$  \\
        $\rho^+\Sigma^0$ & & &  & $0$ & $0$ & $0$ & $-\sqrt{1/2}$   \\
        $\omega\Sigma^+$ & & & &  & $0$ & $0$ & $\sqrt{1/2}$   \\
        $\phi\Sigma^+$ & & & & &  & $0$ & $-1$  \\
        $K^{*+}\Xi^0$ & & & & & & & $1$  \\ \hline
    \end{tabular}
    \hspace{0.5cm}
    \begin{tabular}{ccccccccccccc} \hline
        $Q=0$ & $\bar{K}^{*0} n$ & $K^{*-}p$ & $\rho^0\Lambda$ & $\omega\Lambda$ & $\rho^-\Sigma^+$ & $\rho^0\Sigma^0$ & $\rho^+\Sigma^-$ & $\omega\Sigma^0$ & $\phi\Lambda$ & $K^{*0}\Xi^0$ & $\phi\Sigma^0$ & $K^{*+}\Xi^-$  \\ \hline
        $\bar{K}^{*0} n$ & $2$ & $1$ & $-\sqrt{3}/2$ & $\sqrt{3}/2$ & $0$ & $1/2$ & $1$ & $-1/2$ & $-\sqrt{3/2}$ & $0$ & $1/\sqrt{2}$ & $0$  \\
        $K^{*-}p$ &  & $2$ & $\sqrt{3}/2$ & $\sqrt{3}/2$ & $1$ & $1/2$ & $0$ & $1/2$ & $-\sqrt{3/2}$ & $0$ & $-1/\sqrt{2}$ &  $0$   \\
        $\rho^0\Lambda$ &  &  & $0$ & $0$ & $0$ & $0$ & $0$ & $0$ & $0$ & $-\sqrt{3}/2$ & $0$ &  $\sqrt{3}/2$   \\
        $\omega\Lambda$ &  &  &  & $0$ & $0$ & $0$ & $0$ & $0$ & $0$ & $\sqrt{3}/2$ & $0$ & $\sqrt{3}/2$    \\
        $\rho^-\Sigma^+$ &   &  &  &  & $2$ & $2$ & $0$ & $0$ & $0$ & $1$ & $0$ & $0$   \\
        $\rho^0\Sigma^0$ &  &  &  &  &  & $0$ & $2$ & $0$ & $0$ & $1/2$ & $0$ & $1/2$    \\
        $\rho^+\Sigma^-$ &  &  &  &  &  &  & $2$ & $0$ & $0$ & $0$ & $0$ & $1$    \\
        $\omega\Sigma^0$ &  &  &  &  &  &  &  & $0$ & $0$ & $-1/2$ & $0$ & $1/2$    \\
        $\phi\Lambda$ &  &  &  &  &  &  &  &  & $0$ & $-\sqrt{3/2}$ & $0$ & $-\sqrt{3/2}$    \\
        $K^{*0}\Xi^0$ &  &  &  &  &  &  &  &  &  & $2$ & $1/\sqrt{2}$ & $1$    \\
        $\phi\Sigma^0$ &  &  &  &  &  &  &  &  &  &  & $0$ & $-1/\sqrt{2}$    \\
        $K^{*+}\Xi^-$ &  &  &  &  &  &  &  &  &  &  &  & $2$    \\ \hline
    \end{tabular}
    }
    \caption{$C_{ij}$ coefficients of Eq.~\eqref{eq:interactionkernel} for the $(S=-1,Q=+1)$ and $(S=-1,Q=0)$ sectors, with $C_{ji}=C_{ij}$.}
    \label{tab:Q coefficients}
\end{table*}

The $C_{ij}$ coefficients in Eq.~\eqref{eq:interactionkernel}, in the isospin basis, are taken directly from Ref.~\cite{Oset:2010tof} (Tables 8,9 and 10 in that work) and here are compiled  in  Table \ref{tab:Q coefficients}  for the ($S=-1$, $Q=+1$) and the ($S=-1$, $Q=0$) sectors using the physical basis.

\bibliographystyle{apsrev4-1}
\bibliography{apssamp}

\end{document}